\newcommand{\tran}[1]{}

\newcommand{\conf}[1]{#1} %

\tran{}
\conf{
\documentclass[10pt,preprint]{sigplanconf}
}

\usepackage{url}
\usepackage{alltt} %
\usepackage[T1]{fontenc} %
\usepackage{graphicx}
\usepackage{amsmath}
\usepackage{setspace} %

\newcommand{\Vs}{\vspace{-1.2ex}} 

\newcommand{\myparag}[1]{\vspace{1ex}\noindent{\bf #1.}~}

\newcommand\defn[1]{#1}

\def\mathify#1{\ifmmode{#1}\else\mbox{$#1$}\fi}
\newcommand\m[1]{$#1$} %
\tran{}
\conf{
\newenvironment{code}{\begin{alltt}\small}{\end{alltt}}
\newenvironment{smallcode}{\begin{alltt}\scriptsize}{\end{alltt}\vspace{-1.5ex}}
\newcommand\co[1]{{\small \tt #1}} %
}
\newcommand\p[1]{\m{#1}}
\renewcommand\O[1]{\m{O(#1)}} %
\newcommand{\notes}[1]{} %

\newcommand{\msg}[1]{} %
\newcommand{\tr}[1]{} %
\newcommand{\jnl}[1]{} %

\tran{}%

\begin{document}

\title{From Clarity to Efficiency for Distributed
  Algorithms%
\conf{%
  \thanks{This work was supported in part by NSF under
    grants CCF-1414078, %
    CCF-1248184, %
    CCF-0964196, %
    CNS-0831298, %
    and CCF-0613913; %
    and ONR under grants 
    N000141512208, %
    N000140910651 %
    and N000140710928. %
    \vspace{-29ex} %
}
}
}

\conf{
\authorinfo{\vspace{-1ex}
Yanhong A. Liu \and Scott D. Stoller \and Bo Lin}%
{Computer Science Department, Stony Brook University,
  Stony Brook, NY 11794, USA}%
{\{liu,stoller,bolin\}@cs.stonybrook.edu\vspace{-5ex}}

\maketitle

}%
\tran{}%

\begin{abstract}
  This article describes a very high-level language for clear
  description of distributed algorithms
  and optimizations necessary for generating efficient implementations.
  The language supports high-level control flows where complex
  synchronization conditions can be expressed using high-level
  queries, especially logic quantifications, over message history
  sequences.
  Unfortunately, the programs would be extremely inefficient,
  including consuming unbounded memory, if executed straightforwardly.

  We present new optimizations that automatically transform complex
  synchronization conditions into incremental updates of necessary
  auxiliary values as messages are sent and received.  
  The core of the optimizations is the first general method for
  efficient implementation of logic quantifications.
  We have developed an operational semantics of the language,
  implemented a prototype of the compiler and the optimizations, and
  successfully used the language and implementation on a variety of
  important distributed algorithms.
\notes{}%
\notes{}%
\notes{}%
\end{abstract}

\tran{}%

\conf{

}%


\tran{}

\section{Introduction}

Distributed algorithms are at the core of distributed systems.  Yet,
developing practical implementations of distributed algorithms with
correctness and efficiency assurances remains a challenging, recurring
task.
\tran{}
\begin{itemize}
\item Study of distributed algorithms has relied on either pseudocode
  with English, which is high-level but imprecise, or formal
  specification languages, which are precise but harder to understand,
  lacking mechanisms for building real distributed systems, or not
  executable at all.
\item At the same time, programming of distributed systems has mainly
  been concerned with program efficiency and has relied mostly on the
  use of low-level or complex libraries and to a lesser extent on
  built-in mechanisms in restricted programming models.
\end{itemize}
What's lacking is (1) a simple and powerful language that can express
distributed algorithms at a high level and yet has a clear semantics
for precise execution as well as for verification, and is fully
integrated into widely used programming languages for building real
distributed systems, together with (2) powerful optimizations that can
transform high-level algorithm descriptions into efficient
implementations.

\notes{}

This article describes a very high-level language, DistAlgo, for clear
description of distributed algorithms, combining advantages of
pseudocode, formal specification languages, and programming languages.
\begin{itemize}

\item The main control flow of a process, including sending messages
  and waiting on conditions about received messages, can be stated
  directly as in sequential programs; yield points where message
  handlers execute can be specified explicitly and declaratively.
  
\item Complex synchronization conditions can be expressed using
  high-level queries, especially quantifications, over\jnl{} message history sequences, without manually writing
  message handlers that perform low-level incremental updates and
  obscure control flows.

\notes{}%
\end{itemize}
DistAlgo supports these features by building on an object-oriented
programming language.
We also developed an operational semantics for the language.  The
result is that distributed algorithms can be expressed in DistAlgo
clearly at a high level, %
like in pseudocode, but also precisely, like in formal specification
languages, facilitating formal verification, and can be executed as
part of real applications, as in programming languages.

Unfortunately, programs containing control flows with synchronization
conditions expressed at such a high level are extremely inefficient if
executed straightforwardly: each quantifier can introduce a linear
factor in running time, and any use of the history of messages sent
and received may cause space usage to be unbounded.

We present new optimizations that allow efficient implementations to
be generated automatically, extending previous optimizations to
distributed programs and to the most challenging quantifications.
\begin{itemize}
\item Our method transforms sending and receiving of messages into
  updates to message history sequences, incrementally maintains the
  truth values of synchronization conditions and necessary auxiliary
  values as those sequences are updated, and finally removes those
  sequences as dead code when appropriate.
\item To incrementally maintain the truth values of general
  quantifications, our method first transforms them into aggregations,
  also called aggregate queries.  In general, however, translating
  nested quantifications simply into nested aggregations can incur
  asymptotically more space and time overhead than necessary.  Our
  transformations minimize the nesting of the resulting queries.
\item Quantified order comparisons are used extensively in nontrivial
  distributed algorithms.  They can be incrementalized easily when not
  mixed with other conditions or with each other.  We systematically
  extract single quantified order comparisons and transform them into
  efficient incremental operations.
\end{itemize}
Overall, our method significantly improves time complexities and
reduces the unbounded space used for message history sequences to the
auxiliary space needed for incremental computation.  Systematic
incrementalization also allows the time and space complexity of the
generated programs to be analyzed easily.

\notes{}%

There has been a significant amount of related research, as discussed
in Section~\ref{sec-related}\notes{}.
Our work contains three main contributions:
\begin{itemize}
\item A simple and powerful language for expressing distributed
  algorithms with high-level control flows and synchronization
  conditions, an operational semantics, and full integration into an
  object-oriented language.
\item A systematic method for incrementalizing complex synchronization
  conditions with respect to all sending and receiving of messages in
  distributed programs.
\item A general and systematic method for generating efficient
  implementations of arbitrary logic quantifications together with
  general high-level queries.
\msg{}
\end{itemize}

We have implemented a prototype of the compiler and the optimizations
and experimented with a variety of important distributed algorithms,
including Paxos, Byzantine Paxos, and multi-Paxos.
Our experiments strongly confirm the benefits of the language and the
effectiveness of the optimizations.

This article is a revised version %
of\conf{ Liu et al.~\cite{Liu+12DistPL-OOPSLA}}\tran{}.
The main changes are revised and extended descriptions of the language
and the optimization method, a new formal operational semantics, an
abridged and updated description of the implementation, and a new
description of our experience of using DistAlgo in teaching.

\jnl{}

\section{Expressing distributed algorithms}
\label{sec-dist}

Even when a distributed algorithm appears simple at a high level, it
can be subtle when necessary details are considered, making it
difficult to understand how the algorithm works precisely.  The
difficulty comes from the fact that multiple processes\notes{} must coordinate and synchronize to
achieve global goals, but at the same time, delays, failures, and
attacks can occur.  Even determining the ordering of events is
nontrivial, which is why Lamport's logical clock~\cite{Lam78} is so
fundamental for distributed systems.

\notes{}%

\myparag{Running example}
We use Lamport's distributed mutual exclusion algorithm~\cite{Lam78}
as a running example.  Lamport developed it to illustrate the logical
clock he invented.  The problem is that \m{n} processes access a
shared resource, and need to access it mutually exclusively, in what
is called a critical section (CS), i.e., there can be at most one
process in a critical section at a time.  The processes have no shared
memory, so they must communicate by sending and receiving messages.
Lamport's algorithm assumes that communication channels are reliable
and first-in-first-out (FIFO).

\notes{}%

Figure~\ref{fig-lam-paper} contains Lamport's original description of
the algorithm, except with the notation \m{<} instead of
\m{\longrightarrow} in rule~5 (for comparing pairs of timestamps and
process ids using lexical ordering: \co{(a,b)\,\m{<}\,(a2,b2)} iff
\co{a\,\m{<}\,a2} or \co{a\,\m{=}\,a2} and \co{b\,\m{<}\,b2}) and with
the word ``acknowledgment'' added in rule~5 (for simplicity when
omitting a commonly omitted~\cite{Lynch96,Garg02} small optimization
mentioned in a
footnote). %
This description is the most authoritative, is at a high level, and
uses the most precise English we found.

\begin{figure}[htbp]
  \centering
\conf{
  \small
}
\fbox{
\tran{}
\conf{
\begin{tabular}{@{}p{0.45\textwidth}@{}}
}

\notes{}%

\quad The algorithm is then defined by the following five rules.  For
convenience, the actions defined by each rule are assumed to form a
single event.

\quad 1. To request the resource, process \m{P_i} sends the message
{\it \m{T_m}:\m{P_i} requests resource} to every other process, and puts that
message on its request queue, where \m{T_m} is the timestamp of the
message.

\quad 2. When process \m{P_j} receives the message {\it\m{T_m}:\m{P_i} requests
  resource}, it places it on its request queue and sends a (timestamped)
acknowledgment message to \m{P_i}.

\quad 3. To release the resource, process \m{P_i} removes any
{\it\m{T_m}:\m{P_i} requests resource} message from its request queue and
sends a (timestamped) {\it\m{P_i} releases resource} message to every
other process.

\quad 4. When process \m{P_j} receives a {\it\m{P_i} releases resource}
message, it removes any {\it\m{T_m}:\m{P_i} requests resource} message from
its request queue.

\quad 5. Process \m{P_i} is granted the resource when the following two
conditions are satisfied: (i) There is a {\it\m{T_m}:\m{P_i} requests
  resource} message in its request queue which is ordered before any
other request in its queue by the relation \m{<}. (To define the
relation \m{<} for messages, we identify a message with the event of
sending it.) (ii) \m{P_i} has received an acknowledgment
message from every other process timestamped later than \m{T_m}.

Note that conditions (i) and (ii) of rule 5 are tested locally by
\m{P_i}.
\end{tabular}
}
  \caption{Original description in English.}
  \label{fig-lam-paper}
\end{figure}

The algorithm satisfies safety, liveness, and fairness, and has a
message complexity of \m{3(n-1)}.  It is safe in that at most one
process can be in a critical section at a time.  It is live in that
some process will be in a critical section if there are requests.  It
is fair in that requests are served in the order of the logical
timestamps of the request messages.  Its message complexity is
\m{3(n-1)} in that \m{3(n-1)}
messages are required to serve each request.

\myparag{Challenges}
To understand how this algorithm is carried out precisely, one must
understand how each of the \m{n} processes acts as both \m{P_i} and
\m{P_j} in interactions with all other processes.  Each process must
have an order of handling all the events according to the five rules,
trying to reach its own goal of entering and exiting a critical
section while also responding to messages from other processes.  It
must also keep testing the complex condition in rule~5 as events
happen.

State machine based formal specifications have been used to fill in
such details precisely, but at the same time, they are lower-level and
harder to understand.  For example, a formal specification of
Lamport's algorithm in I/O automata~\cite[pages 647-648]{Lynch96}
occupies about one and a fifth pages, most of which is double-column.

To actually implement distributed algorithms, details for many
additional aspects must be added, for example, creating processes,
letting them establish communication channels with each other,
incorporating appropriate logical clocks (e.g., Lamport clock or
vector clock~\cite{Mattern89}) if needed, guaranteeing the specified
channel properties (e.g., reliable, FIFO), and integrating the
algorithm with the application (e.g., specifying critical section
tasks and invoking the code for the algorithm as part of the overall
application).  Furthermore, how to do all of these in an easy and
modular fashion?
\notes{}%

\notes{}%

\myparag{Our approach}
We address these challenges with the DistAlgo language, compilation to
executable programs, and especially optimization by incrementalization
of expensive synchronizations, described in Sections~\ref{sec-lang},
\ref{sec-compile}, and~\ref{sec-inc}, respectively.  An unexpected
result is that incrementalization led us to discover simplifications
of Lamport's original algorithm in Figure~\ref{fig-lam-paper}; the
simplified algorithm can be expressed using basically two \co{send}
statements, a \co{receive} definition, and an \co{await} statement.

The results on the running example are shown in
Figures~\ref{fig-lam-orig}--\ref{fig-lam-simp}, with details explained
later.
Figure~\ref{fig-lam-orig} shows Lamport's original algorithm expressed
in DistAlgo; it also includes configuration and setup for running 
50 processes each trying to enter critical section at some point
during its execution.
Figures~\ref{fig-lam-inc} and~\ref{fig-lam-inc-min} show two
alternative optimized programs after incrementalization; all lines
with comments are new except that the \co{await} statement is
simplified.
Figure~\ref{fig-lam-simp} shows the simplified algorithm.

\msg{}

\section{DistAlgo Language}
\label{sec-lang}

To support distributed programming at a high level, four main concepts
can be added
to commonly used high-level programming languages, 
especially object-oriented languages, such as Python and Java:
(1) distributed processes, and sending messages,
(2) control flows with yield points and waits, and receiving
messages,
(3) synchronization conditions using high-level queries of message
history sequences, and
(4) configuration of processes and communication mechanisms.
DistAlgo supports these concepts, with options and generalizations for
ease of programming, as described below.
A formal operational semantics for DistAlgo is presented in Appendix
A.

\myparag{Processes and sending of messages}
Distributed processes are concurrent executions of programmed
instructions, like threads in Java and Python, except that each
process has its private memory, not shared with other processes, and
processes communicate by message passing.  Three main constructs are
used, for defining processes, creating processes, and sending
messages.

A process definition is of form (1) below.  It defines a type
\p{p} of processes, by defining a class \p{p} that extends class
\co{process}.  The \co{\p{process\_body}} is a set of method definitions
and handler definitions, to be described.
\begin{code}
    class \p{p} extends process:\Vs
\hfill{\normalfont (1)}\Vs
      \p{process\_body}
\end{code}
A special method \co{setup} may
be defined in \co{\p{process\_body}} for
initially setting up data in the process before the process's
execution starts.
A special method \co{run()}
may 
be defined in \co{\p{process\_body}} for carrying out the main
flow of execution.
A special variable \co{self} refers to the process itself.

A process creation statement is of form (2) below.  It creates \p{n}
new processes of type \p{p} at each node in the value of expression
\co{\p{node\_exp}}, and returns\notes{} the resulting process
or set of processes\notes{}.  A node is a running
DistAlgo program on a machine, and is identified by the host name of
the machine plus the name of the running DistAlgo program that can be
specified when starting the program.
\begin{code}
    \p{n} new \p{p} at \p{node\_exp}  \hfill{\normalfont (2)}
\end{code}
The number \p{n} and the \co{at} clause are optional; the defaults are 1 and
the local node, respectively.
A new process can be set up by calling its \co{setup} method. 
A call
\co{start()} on the process then starts the execution of its \co{run()}
method.

A statement for sending messages is of form (3) below.  It sends
the message that is the value of expression \p{mexp} to the process or
set of processes that is the value of expression \p{pexp}.
\begin{code}
    send \p{mexp} to \p{pexp}  \hfill{\normalfont (3)}
\end{code}
A message can be any value but is by convention a tuple whose first
component is a string, called a tag, indicating the kind of the
message.

\myparag{Control flows and handling of received messages}
The key idea is to use labels to specify program points where control
flow can yield to handling of messages and resume afterwards.  Three
main constructs are used, for specifying yield points, handling of
received messages, and synchronization.

\notes{}
A yield point preceding a statement is of form (4) below, where
identifier \p{l} is a label.  It specifies that point in the program
as a place where control\notes{} yields to handling of un-handled messages, if any, and
resumes afterwards.
\begin{code}
    -- \p{l}  \hfill{\normalfont (4)}
\end{code}
The label \co{l} is optional; it can be omitted when this yield point
is not explicitly referred to in any handler definitions, defined
next.

A handler definition, also called a \co{receive} definition, is of 
form (5) below.
It handles, at yield points labeled \p{l_1}, ..., \p{l_j}, un-handled
messages that match some \p{mexp_i} sent\notes{}
from\notes{} \p{pexp_i}, where \p{mexp_i} and
\p{pexp_i} are parts of a tuple pattern;
previously unbound variables in a pattern are bound to the
corresponding components in the value matched.  The
\co{\p{handler\_body}} is a sequence of statements to be executed for
the matched messages.
\begin{code}
    receive \p{mexp\sb{1}}\mbox{\!} from \p{pexp\sb{1}}, ..., \p{mexp\sb{k}} from \p{pexp\sb{k}}
        at \p{l\sb{1}}, ..., \p{l\sb{j}}:  \hfill{\normalfont (5)}
      \p{handler\_body}
\end{code}
The \co{from} and \co{at} clauses are optional; the defaults are any
process and all yield points, respectively.  
If the \co{from} clause is used, each message is automatically extended
with the process id of the sender.
A tuple pattern is a tuple in which each component is a non-variable
expression, a variable possibly prefixed with "\co{=}", a wildcard, or
recursively a tuple pattern.  A non-variable expression or a variable
prefixed with ``\co{=}'' means that the corresponding component of the
tuple being matched must equal the value of the non-variable
expression or the variable, respectively, for pattern matching to
succeed.  A variable not prefixed with ``\co{=}'' matches any value and
becomes bound to the corresponding component of the tuple being
matched.
A wildcard, written as ``\co{\_}'', matches any value.
Support for \co{receive} mimics common usage in pseudocode, allowing a
message handler to be associated with multiple yield points without
using method definition and invocations.
As syntactic sugar, a \co{receive} that is handled at only one yield
point can be written at that point.
\notes{}%
\notes{}

Synchronization and associated actions can be expressed using general,
non\-deterministic \co{await} statements.  
A simple \co{await} statement is one of the two forms in (6) below.
It waits for the value of Boolean-valued expression \p{bexp} to become
true, for the first form, or waits for a timeout after time period
\p{t}, for the second form.
\begin{code}
    await \p{bexp}\Vs
\hfill{\normalfont (6)}\Vs
    await timeout \p{t}
\end{code}
A general, nondeterministic \co{await} statement is of form (7) below.
It waits for any of the values of expressions \p{bexp\sb{1}},
..., \p{bexp\sb{k}} to become true or a timeout after time period \p{t},
and then nondeterministically selects one\notes{} of
statements \p{stmt\sb{1}}, ..., \p{stmt\sb{k}}, \p{stmt} whose
corresponding conditions are satisfied to execute.  The \co{or} and
\co{timeout} clauses are optional.
\begin{code}
    await \p{bexp\sb{1}}: \p{stmt\sb{1}}
    or ...\Vs
\hfill{\normalfont (7)}\Vs
    or \p{bexp\sb{k}}: \p{stmt\sb{k}}
    timeout \p{t}: \p{stmt}
\end{code}
An \co{await} statement must be preceded by a yield point, for
handling messages while waiting; if a yield point is not specified
explicitly, the default is that all message handlers can be executed
at this point.

These few constructs make it easy to specify any process that has its
own flow of control while also responding to messages.  It is also
easy to specify any process that only responds to messages, for
example, by writing just \co{receive} definitions and a \co{run()}
method containing only \co{await false}.

\myparag{Synchronization conditions using high-level queries}
Synchronization conditions and other conditions can be expressed using
high-level queries---quantifications, comprehensions, and\tran{}
aggregations---over sets of processes and sequences of messages.
High-level queries
are used commonly in distributed algorithms because (1) they make
complex synchronization conditions clearer and easier to write, and
(2) the complexity\notes{} of
distributed algorithms is measured by round complexity and message
complexity, not time complexity of local processing.

Quantifications are especially common because they directly capture
the truth values of synchronization conditions.  We discovered a
number of errors in our initial programs that\notes{} were
written using aggregations in place of quantifications before we
developed the method to systematically optimize quantifications.  For
example, we regularly expressed ``\co{v} is larger than all elements of
\co{s}'' as \co{v~>~max\,s} and either forgot to handle the case that
\co{s} is empty or handled it in an ad~hoc fashion.  Naive use of
aggregation operators like \co{max} may also hinder generation of more
efficient implementations.

We define operations on sets; operations on sequences are the same
except that elements are processed in order, and square brackets are
used in place of curly braces.
\begin{itemize}
\item A quantification is a query of one of the two forms in (8) below,
  called existential and universal quantifications, respectively, plus
  a set of parameters---variables whose values are bound before the
  query.
  For a query to be well-formed, every variable in it must be
  \defn{reachable} from a parameter---be a parameter or recursively be
  the left-side variable of a membership clause whose right-side
  variables are reachable.
  Given values of parameters, the query returns \co{true} iff for some
  or all, respectively, combinations of values of variables that
  satisfy all membership clauses \co{\p{v\sb{i}} in \p{sexp\sb{i}}},
  expression \p{bexp} evaluates to \co{true}.  When an existential
  quantification returns \co{true}, all variables in the query are also
  bound to a combination of values, called a witness, that satisfy all
  the membership clauses and condition \p{bexp}.
\begin{code}
    some \p{v\sb{1}} in \p{sexp\sb{1}}, ..., \p{v\sb{k}} in \p{sexp\sb{k}} | \p{bexp}\Vs
\hfill{\normalfont (8)}\Vs
    each \p{v\sb{1}} in \p{sexp\sb{1}}, ..., \p{v\sb{k}} in \p{sexp\sb{k}} | \p{bexp}
\end{code}
For example, the following query returns \co{true} iff each element in
\co{s} is greater than each element in \co{s2}.
\begin{code}
    each x in s, x2 in s2 | x > x2
\end{code}
For another example, the following query, containing a nested
quantification, returns \co{true} iff some element in \co{s} is greater
than each element in \co{s2}.  Additionally, when the query returns
true, variable \co{x} is bound to a witness---an element in \co{s} that
is greater than each element in \co{s2}.
\begin{code}
    some x in s | each x2 in s2 | x > x2
\end{code}

\item A comprehension is a query of form (9) below. 
  Given values of parameters, the query returns the set of values of
  \p{exp} for all combinations of values of variables that satisfy all
  membership clauses \co{\p{v\sb{i}} in \p{sexp\sb{i}}} and condition
  \p{bexp}.
\begin{code}
    \{\,\p{exp}: \p{v\sb{1}} in \p{sexp\sb{1}}, ..., \p{v\sb{k}} in \p{sexp\sb{k}}| \p{bexp}\,\}  \hfill{\normalfont (9)}
\end{code}
For example, the following query returns the set of products of \co{x}
in \co{s} and \co{x2} in \co{s2} where \co{x} is greater than \co{x2}.
\begin{code}
    \{x*y: x in s, x2 in s2 | x > x2\}
\end{code}
We abbreviate \co{\{\p{v}: \p{v} in \p{sexp} | \p{bexp}\}} as \co{\{\p{v} in
  \p{sexp} | \p{bexp}\}}.

\item An aggregation, also called an aggregate query, is a query of
  one of the two forms in (10) below, where \co{\p{agg}} is an
  aggregation operator, including \co{count},
 \co{sum}, \co{min}, and \co{max}.
 Given values of parameters, the query returns the value of applying
 \co{\p{agg}} to the set value of \co{\p{sexp}}, for the first form, or
 to the multiset of values of \p{exp} for all combinations of values
 of variables that satisfy all membership clauses \co{\p{v\sb{i}} in
   \p{sexp\sb{i}}} and condition \p{bexp}, for the second form.
\begin{code}
    \p{agg} \p{sexp}\Vs
\hfill{\normalfont (10)}\Vs
    \p{agg} \{\,\p{exp}: \p{v\sb{1}} in \p{sexp\sb{1}}, ..., \p{v\sb{k}} in \p{sexp\sb{k}}| \p{bexp}\,\}
\end{code}
  
\item In the query forms above, each \p{v_i} can also be a tuple
  pattern \p{t_i}.  
  Variables in \p{t_i} are bound to the
  corresponding components in the matched elements of the value of
  \p{sexp\sb{i}}.
  We omit \co{|\,\p{bexp}} when \p{bexp} is \co{true}.
\jnl{}

\end{itemize}
We use \co{\{\}}\notes{} for empty set; use \co{s.add(x)} and
\co{s.del(x)} for element addition and deletion, respectively; and use
\co{x in s} and \co{x not in s} for membership test and its negation,
respectively.
We assume that hashing is used in implementing sets, and the expected
time of
set initialization, element addition and removal, and membership test
is \O{1}.
We consider operations that involve iterations over sets and sequences
to be expensive; each iteration over a set or sequence incurs a cost
that is linear in the size of the set or sequence.  All
quantifications, comprehensions, and aggregations are considered
expensive.

DistAlgo has built-in sequences \co{received} and \co{sent}, containing all
messages received and sent, respectively, by a process.
\begin{itemize}

\item Sequence \co{received} is updated only at yield points; 
after a message arrives, it will be handled when execution reaches 
the next yield point, by adding the message to \co{received} 
and running matching \co{receive} definitions, if any,
associated with the yield point.
  We use \co{received m from p} interchangeably with \co{m from p in
    received} to mean that message \co{m} from process \co{p} is in \co{received};
  \co{from p} is optional, but when specified, each message in
  \co{received} is automatically extended with the process id of the
  sender.

\item Sequence \co{sent} is updated at each \co{send} statement; each
  message sent to a process is added to \co{sent}.
  We use \co{sent m to p} interchangeably with \co{m to p in
    sent} to mean that message \co{m} to process \co{p} is in \co{sent};
  \co{to p} is optional, but when specified, \co{p} is the
  process to which \co{m} was sent as specified in the %
  \co{send} statement.

\end{itemize}
If implemented straightforwardly, \co{received} and \co{sent} can create
a huge memory leak, because they can grow unboundedly, preventing
their use in practical programming.  Our method can remove them by
maintaining only auxiliary values that are needed for incremental
computation.

\myparag{Configuration}
One can specify channel types, handling of messages, 
and other configuration items.  Such specifications are declarative,
so that algorithms can be expressed without unnecessary implementation
details.
We describe a few basic kinds of configuration items.

First, one can specify the types of channels for passing messages.
For example, the following statement configures all channels to be
FIFO.
\begin{code}
    configure channel = fifo
\end{code}
Other options for \co{channel} include \co{reliable} and\conf{\linebreak}
\,\co{\{reliable,~fifo\}}.\, When either \co{fifo} or \co{reliable} is
included, TCP is used for process communication; otherwise, UDP is
used.
In general, channels can also be configured separately for messages
from any set of processes to any set of processes.

One can specify how much effort is spent processing messages at yield
points.  For example,
\begin{code}
    configure handling = all
\end{code}
configures the system to handle all %
un-handled messages at each yield point; 
this is the default.
For another example, one can specify a time limit.  One can also
specify different handling effort for different yield points.

Logical clocks~\cite{Lam78,Fidge88,Mattern89} are used in many
distributed algorithms.  One can specify the logical clock, e.g.,
Lamport clock, that is used:
\begin{code}
    configure clock = Lamport
\end{code}
It configures sending and receiving of messages to update the clock
appropriately.  A call\tran{} \co{logical\_time()} returns the
current value of the logical clock.
\notes{}

Overall, a DistAlgo program consists of a set of process definitions, a
method \co{main}, and possibly other, conventional program parts.  Method
\co{main} specifies the configurations and creates, sets up, and starts a
set of processes.
DistAlgo language constructs can be used in process definitions and method
\co{main} and are implemented according to the semantics described; other,
conventional program parts are implemented according to their conventional
semantics.

\notes{}%

\myparag{Other language constructs}
For other constructs, we use those in high-level object-oriented
languages.  We mostly use Python syntax (indentation for scoping,
'\co{:}' for elaboration, '\co{\#}' for comments, etc.) for
succinctness, except with \co{\p{v}\,:=\,\p{exp}} for assignment and
with a few conventions from Java (keyword \co{extends} for subclass,
keyword \co{new} for object creation, and omission of \co{self}, the
equivalent of \co{this} in Java, when there is no ambiguity\notes{}) for ease of reading.

\myparag{Example} 
Figure~\ref{fig-lam-orig} shows Lamport's algorithm expressed in
DistAlgo.  The algorithm in Figure~\ref{fig-lam-paper} corresponds to
the body of \co{mutex} and the two \co{receive} definitions, 16 lines total;
the rest of the program, 14 lines total, shows how the algorithm is
used in an application.  The execution of the application starts with
method \co{main}, which configures the system to run (lines 25-30).
Method \co{mutex} and the two \co{receive} definitions are executed when
needed and follow the five rules in Figure~\ref{fig-lam-paper} (lines
5-21). 
Recall that there is an implicit yield point before the \co{await}
statement.

Note that Figure~\ref{fig-lam-orig} is not meant to replace
Figure~\ref{fig-lam-paper}, but to realize Figure~\ref{fig-lam-paper}
in a precisely executable manner.  Figure~\ref{fig-lam-orig} 
is meant to be high-level, compared with lower-level specifications 
and programs.

\begin{figure}[htbp]
\begin{smallcode}
\tran{} 1 class P extends process:
\tran{} 2   def setup(s):
\tran{} 3     self.s := s                # set of all other processes
\tran{} 4     self.q := \{\}               # set of pending requests
\tran{}   
\tran{} 5   def mutex(task):             # run task with mutual exclusion
\tran{} 6     -- request
\tran{} 7     self.t := logical_time()                        # 1 in Fig 1
\tran{} 8     send ('request', t, self) to s                  #
\tran{} 9     q.add(('request', t, self))                     #
\tran{}                                  # wait for own req < others in q
\tran{}                                  #  and for acks from all in s
\tran{}10     await each ('request', t2, p2) in q |           # 5 in Fig 1
\tran{}               (t2,p2) != (t,self) implies (t,self) < (t2,p2)
\tran{}11           and each p2 in s |                        #
\tran{}                some received('ack', t2, =p2) | t2 > t
\tran{}12     task()                     # critical section
\tran{}13     -- release
\tran{}14     q.del(('request', t, self))                     # 3 in Fig 1 
\tran{}15     send ('release', logical_time(), self) to s     #  
\tran{}
\tran{}16   receive ('request', t2, p2):                      # 2 in Fig 1
\tran{}17     q.add(('request', t2, p2))                      #
\tran{}18     send ('ack', logical_time(), self) to p2        #
\tran{}
\tran{}19   receive ('release', _, p2):                       # 4 in Fig 1
\tran{}20     for ('request', t2, =p2) in q:                  #
\tran{}21       q.del(('request', t2, p2))                    #
\tran{}
\tran{}22   def run():                   # main method for the process
\tran{}       ...                        # do non-CS tasks of the process
\tran{}23     def task(): ...            # define critical section task
\tran{}24     mutex(task)                # run task with mutual exclusion
\tran{}       ...                        # do non-CS tasks of the process
\tran{}                                  
\tran{}25 def main():                    # main method for the application
\tran{}     ...                          # do other tasks of the application
\tran{}26   configure channel = \{reliable, fifo\}      
\tran{}                                  # use reliable and FIFO channel
\tran{}27   configure clock = Lamport    # use Lamport clock
\tran{}28   ps := 50 new P               # create 50 processes of P class
\tran{}29   for p in ps: p.setup(ps-\{p\}) # pass to each process other processes
\tran{}30   for p in ps: p.start()       # start the run method of each process
\tran{}     ...                          # do other tasks of the application
\end{smallcode}
  \caption{Original algorithm (lines 6-21) in a complete program in DistAlgo.}
  \label{fig-lam-orig}
\end{figure}

\jnl{}

\section{Compiling to executable programs}
\label{sec-compile}

Compilation generates code to create processes on the specified
machine, take care of sending and receiving messages, and realize the
specified configuration.  In particular, it inserts appropriate
message handlers at each yield point.

\myparag{Processes and sending of messages}
\notes{}%
Process creation is compiled to creating a process on the specified or
default machine and that has a private memory space for its
fields.\notes{}
Each process is implemented using two threads: a main thread that
executes the main flow of control of the process, and a helper thread
that receives and enqueues messages sent to this process.
Constructs involving a set of processes, such as \co{n new P},
can easily be compiled into loops.
\notes{}%

Sending a message \co{m} to a process \co{p} is compiled into calls to a
standard message passing API.  If the sequence \co{sent} is used in the
program, we also insert \co{sent.add(m to p)}.
Calling a method on a remote process object is compiled into a remote
method call.
\notes{}%

\myparag{Control flows and handling of received messages}
Each yield point \co{l} is compiled into a call to a message handler
method \co{l()} that updates the sequence \co{received}, if
\co{received} is used in the program, and executes the bodies of the
\co{receive} definitions whose \co{at} clause includes \co{l}.
Precisely:
\begin{itemize}

\item Each \co{receive} definition is compiled into a method that takes
  a message \co{m} as argument, matches \co{m} against the message
  patterns in the \co{receive} clause, and if the matching succeeds,
  binds the variables in the %
  matched pattern appropriately, and executes the statement in the
  body of this \co{receive} definition.

\item Method \co{l()} compiled for yield point \co{l} does the
  following: for each un-handled message \co{m} from \co{p} to be
  handled, (1) execute \co{received.add(m from p)} if \co{received} is
  used in the program, (2)
  call the methods generated from the \co{receive}
  definitions %
  whose \co{at} clause includes \co{l}, and (3) remove \co{m} from the
  message queue.

\end{itemize}

\notes{}%

An \co{await} statement can be compiled into a synchronization using
busy-waiting or blocking.  
We use blocking to wait until a new message arrives or the timeout
specified in \co{await} is reached.

\notes{}%

\myparag{Configuration}
Configuration options are taken into account during compilation in a
straightforward way.  Libraries and modules are used as much as
possible.  For example, when \co{fifo} or \co{reliable} channel is
specified, the compiler can generate code that uses TCP sockets.
\notes{}%
\notes{}%

\tran{}
\conf{
\section{$\!\!\!$Incrementalizing\,expensive\,synchronizations$\!$}
}
\label{sec-inc}

Incrementalization transforms expensive computations into efficient
incremental computations with respect to updates to the values on
which the computations depend.  It (1) identifies all expensive
queries, (2) determines all updates that may affect the query result,
and (3) transforms the queries and updates into efficient
incremental computations.  Much of incrementalization has been studied
previously, as discussed in Section~\ref{sec-related}.

The new method here is for (1) systematic handling of quantifications
for synchronization as expensive queries, especially nested
alternating universal and existential quantifications and
quantifications containing complex order comparisons and (2)
systematic handling of updates caused by all sending, receiving, and
handling of messages in the same way as other updates in the program.
The result is a drastic reduction of both time and space complexities.

\myparag{Expensive computations using quantifications}
Expensive computations in general involve repetition, including loops,
recursive functions, comprehensions, aggregations, and quantifications
over collections.  Optimizations were studied most for loops, less for
recursive functions, comprehensions, and aggregations, and least for
quantifications, basically corresponding to how frequently these
constructs have traditionally been used in programming.  However,
high-level queries are increasingly used in programming, and
quantifications are dominantly used in writing synchronization
conditions and assertions in specifications and very high-level
programs.
Unfortunately, if implemented straightforwardly, each quantification
introduces a cost factor that is linear in the size of the collection
quantified over.

Optimizing expensive quantifications in general is difficult, which is
a main reason that they are not used in practical programs, not even
logic programs, and programmers manually write more complex and
error-prone code.  The difficulty comes from expensive enumerations
over collections and complex combinations of join conditions.  We
address this challenge by converting quantifications into aggregations
that can be optimized systematically using previously studied methods.
However, a quantification can be converted into multiple forms of
aggregations.  Which one to use depends on what kinds of updates
must be handled, and on how the query can be incrementalized under
those updates.  Direct conversion of nested quantifications into
nested aggregations can lead to much more complex incremental
computation code and asymptotically worse time and space complexities
for maintaining the intermediate query results.

Note that, for an existential quantification, we convert it to a more
efficient aggregation if a witness is not needed; if a witness is
needed, we incrementally compute the set of witnesses.

\myparag{Converting quantifications to aggregations}
\conf{$\!\!$}We present all converted forms here and describe which forms
to use after we discuss the updates that must be handled.  
The correctness of all rules presented have been proved, manually, using
first-order logic and set theory.  These rules ensure that the value
of a resulting query expression equals the value of the original
quantified expression.

Table~\ref{tab-quant} shows general rules for converting single
quantifications into equivalent aggregations that use aggregation
operator \co{count}.
For converting universal quantifications, either rule 2 or 3 could be
used.
The choice does not affect the asymptotic cost, but only small
constant factors: rule 2 requires maintaining \co{count s}, and rule 3
requires computing \co{not}; the latter is generally faster unless
\co{count s} is already needed for other purposes, and is certainly
faster when \co{not bexp} can be simplified, e.g., when \co{bexp} is a
negation.
The rules in Table~\ref{tab-quant} are general because \co{bexp} can be
any Boolean expression, but they are for converting single
quantifications.  Nested quantifications can be converted one at a
time from inside out, but the results may be much more complicated
than necessary.
For example,
\begin{code}
    \co{each x in s\,|\,some x2 in s2\,|\,bexp}   
\end{code}
would be converted using rule~1 to 
\begin{code}
    \co{each x in s\,|\,count \{x2 in s2\,|\,bexp\} != 0}
\end{code}%
and then using rule~2 to\vspace{-0ex}
\begin{code}
    count \{x in s\,|\,count \{x2 in s2\,|\,bexp\} != 0\} \conf{
    }= count s 
\end{code}
A simpler conversion is possible for this example, using a rule in
Table~\ref{tab-nest}, described next.

\notes{}%
\notes{}%
\begin{table}[htbp]
\conf{
  \small
}
  \centering
  \caption{Rules for converting single quantifications.}
\begin{tabular}{@{}c@{}|@{\tran{}~}l@{~\hfill}|@{}l@{}}\hline
   & \hspace{3ex}Quantification           & \hspace{12ex}Aggregation\\\hline
  ~\,1~\,& 
  \co{some x in s\,|\,bexp} &
  \begin{tabular}{@{\tran{}~}l@{}}
    \co{count \{x in s\,|\,bexp\} != 0} %
  \end{tabular}\\\hline
  \begin{tabular}{@{~\,}c@{~\,}}
    2\\\hline 3
  \end{tabular} &
  \co{each x in s\,|\,bexp} & 
  \begin{tabular}{@{\tran{}~}l@{}}
    \co{count \{x in s\,|\,bexp\} = count s}\\\hline
    \co{count \{x in s\,|\,not bexp\} = 0}
  \end{tabular}\\\hline
\end{tabular}

  \label{tab-quant}
\end{table}

Table~\ref{tab-nest} shows general rules for converting nested
quantifications into equivalent, but non-nested, aggregations that use
aggregation operator \co{count}.  These rules yield much simpler results
than repeated use of the rules in Table~\ref{tab-quant}.  For example,
rule~2 in this table yields a much simpler result than using two rules
in Table~\ref{tab-quant} in the previous example.  More significantly,
rules~1, 4, and 5 generalize to any number of the same quantifier, and
rules~2 and 3 generalize to any number of quantifiers with one
alternation.
\notes{}%
We have not encountered more complicated quantifications than these in the
algorithms we found.  It is well known that more than one alternation is
rarely used, so commonly used quantifications can all be converted to
non-nested aggregations.
For example, in twelve different algorithms expressed in
DistAlgo~\cite{Liu+12DistPL-OOPSLA}, there are a total of 50
quantifications but no occurrence of more than one alternation.
\tran{}
\conf{\begin{table*}[htbp]}
  \small
  \centering
  \caption{Rules for converting nested quantifications.}
{
\begin{tabular}{@{}c@{}|@{~}l@{~\hfill}|@{}l@{}}\hline
   & \hspace{9ex}Nested Quantifications & \hspace{23ex}Aggregation\\\hline
  1& \co{some x in s | some x2 in s2 | bexp} & ~~\co{count \{(x,x2):~x in s, x2 in s2 | bexp\} != 0}\\\hline
  2& \co{each x in s | some x2 in s2 | bexp} & ~~\co{count \{x:~x in s, x2 in s2 | bexp\} = count s}\\\hline
  3& \co{some x in s | each x2 in s2 | bexp} & ~~\co{count \{x:~x in s, x2 in s2 | not bexp\} != count s}\\\hline
  \begin{tabular}{@{\,~}c@{\,~}}
    4\tran{}
    \\\hline
    5
  \end{tabular} &
  \co{each x in s | each x2 in s2 | bexp} & 
  \begin{tabular}{@{~~}l@{~\hfill}}
    \co{count \{(x,x2):~x in s, x2 in s2 | bexp\} = }\\
    \co{count \{(x,x2):~x in s, x2 in s2\}}\\\hline
    \co{count \{(x,x2):~x in s, x2 in s2 | not bexp\} = 0}
  \end{tabular}\\\hline
\end{tabular}
}

  \label{tab-nest}
\tran{}
\conf{\end{table*}}

Table~\ref{tab-cmp} shows general rules for converting single
quantifications with a single order comparison, for any linear order,
into equivalent queries that use aggregation operators \co{max} and
\co{min}.
\notes{}%
These rules are useful because \co{max} and \co{min} can in general be
maintained incrementally in \O{\log n} time %
with \O{n} space overhead.  Additionally, when there are only element
additions, \co{max} and \co{min} can be maintained most efficiently in
\O{1} time and space.
\begin{table}[htbp]
\conf{
  \small
}
  \centering
  \caption{Rules for single quantified order comparison.}
{
\tran{}
\begin{tabular}{@{~}c@{}|@{}l@{}|@{~\,}l@{~}}\hline
  & \hspace{7.5ex}Existential & \hspace{6ex}Aggregation\\\hline
  \begin{tabular}{@{~~~}c@{~~~}}
    1\\\hline 2
  \end{tabular} &
  \begin{tabular}{l}
    \co{some x in s | y <= x}\\\hline
    \co{some x in s | x >= y}
  \end{tabular}
  &\co{s != \{\} and} \co{y <= max s}\\\hline
  \begin{tabular}{@{~~~}c@{~~~}}
    3\\\hline 4
  \end{tabular} &
  \begin{tabular}{l}
    \co{some x in s | y >= x}\\\hline
    \co{some x in s | x <= y} 
  \end{tabular}
  &\co{s != \{\} and} \co{y >= min s}\\\hline
  \begin{tabular}{@{~~~}c@{~~~}}
    5\\\hline 6
  \end{tabular} &
  \begin{tabular}{l@{~~~\,\,\,\,\,}}
    \co{some x in s | y < x}\\\hline
    \co{some x in s | x > y}
  \end{tabular}
  &\co{s != \{\} and} \co{y < max s}~\,\\\hline
  \begin{tabular}{@{~~~}c@{~~~}}
    7\\\hline 8
  \end{tabular} &
  \begin{tabular}{l@{~~~\,\,\,\,\,}}
    \co{some x in s | y > x}\\\hline
    \co{some x in s | x < y}
  \end{tabular}
  &\co{s != \{\} and} \co{y > min s}~\,\\\hline

\notes{}%
\end{tabular}\medskip
\tran{}

\begin{tabular}{@{~}c@{}|@{}l@{}|@{~\,}l@{\hspace{2.8ex}}}\hline
    & \hspace{7.5ex}Universal & \hspace{6ex}Aggregation\\\hline
  \begin{tabular}{@{}c@{~~}}
    9\\\hline 10
  \end{tabular} &
  \begin{tabular}{l}
    \co{each x in s | y <= x}\\\hline
    \co{each x in s | x >= y}
  \end{tabular}
  &
  \begin{tabular}{@{}l@{}}
    \co{s = \{\} or} \co{y <= min s}
  \end{tabular}\\\hline
  \begin{tabular}{@{}c@{~~}}
    11\\\hline 12
  \end{tabular} &
  \begin{tabular}{l}
    \co{each x in s | y >= x}\\\hline
    \co{each x in s | x <= y} 
  \end{tabular}
  & \co{s = \{\} or} \co{y >= max s}\\\hline
  \begin{tabular}{@{}c@{~~}}
    13\\\hline 14
  \end{tabular} &
  \begin{tabular}{l@{~~~\,\,\,\,\,}}
    \co{each x in s | y < x}\\\hline
    \co{each x in s | x > y} 
  \end{tabular}
  & \co{s = \{\} or} \co{y < min s}~\,\\\hline
  \begin{tabular}{@{}c@{~~}}
    15\\\hline 16
  \end{tabular} &
  \begin{tabular}{l@{~~~\,\,\,\,\,}}
    \co{each x in s | y > x}\\\hline
    \co{each x in s | x < y}
  \end{tabular}
  & \co{s = \{\} or} \co{y > max s}~\,\\\hline

\notes{}%
\end{tabular}
\tran{}
}

  \label{tab-cmp}
\end{table}

Table~\ref{tab-extract} shows general rules for decomposing Boolean
combinations of conditions in quantifications,
to obtain quantifications with simpler conditions.
In particular, Boolean combinations of order comparisons and other
conditions can be transformed to extract quantifications each with a
single order comparison, so the rules in Table~\ref{tab-cmp} can be
applied, and Boolean combinations of inner quantifications and other
conditions can be transformed to extract directly nested
quantifications, so the rules in Table~\ref{tab-nest} can be applied.
For example,
\begin{code}
    \co{each x in s\,|\,bexp implies y < x}
\end{code}
can be converted using rule~8 in Table~\ref{tab-extract} to
\begin{code}
    \co{each x in \{x in s\,|\,bexp\}\,|\,y < x}   
\end{code}
which can then be converted using rule~13 of Table~\ref{tab-cmp}
to\vspace{-0ex}
\begin{code}
    \co{\{x in s\,|\,bexp\} = \{\} or y < min \{x in s\,|\,bexp\}}
\end{code}
\notes{}
\notes{}%
\notes{}%
\begin{table}[htbp]
\conf{
  \small
}
  \centering
  \caption{Rules for decomposing conditions to extract quantified comparisons.}
{
\begin{tabular}{l|l@{~\hfill}|l@{~\hfill}}\hline
 & \hspace{\conf{2}\tran{}ex}Quantification & \hspace{\conf{3}\tran{}ex}Decomposed Quantifications\\\hline
1&
\begin{tabular}{@{}l@{}}
  \co{some x in s} \conf{\\}
  \co{| not e}  
\end{tabular}
 &
 \begin{tabular}{@{}l@{}}
   \co{not each x in s | e} \conf{\\}
 \end{tabular}\\\hline
2&
\begin{tabular}{@{}l@{}}
  \co{some x in s} \conf{\\}
  \co{| e1 and e2} 
\end{tabular}
 &
 \begin{tabular}{@{}l@{}}
   \co{some x in \{x in s | e1\}} \conf{\\}
   \co{| e2}
 \end{tabular}\\\hline
3&
\begin{tabular}{@{}l@{}}
  \co{some x in s}  \conf{\\}
  \co{| e1 or e2}
\end{tabular}
 &
 \begin{tabular}{@{}l@{}}
   \co{(some x in s | e1) or} \conf{\\}
   \co{(some x in s | e2)}
 \end{tabular}\\\hline
4&
\begin{tabular}{@{}l@{}}
  \co{some x in s}  \conf{\\}
  \co{| e1 implies e2}
\end{tabular}
 &
 \begin{tabular}{@{}l@{}}
   \co{(some x in s | not e1) or} \conf{\\}
   \co{(some x in s | e2)}   
 \end{tabular}\\\hline

5&
\begin{tabular}{@{}l@{}}
  \co{each x in s}  \conf{\\}
  \co{| not e}  

\end{tabular}
 &
 \begin{tabular}{@{}l@{}}
   \co{not some x in s | e} \conf{\\}
 \end{tabular}\\\hline
6&
\begin{tabular}{@{}l@{}}
  \co{each x in s}  \conf{\\}
  \co{| e1 and e2}
\end{tabular}
 &
 \begin{tabular}{@{}l@{}}
   \co{(each x in s | e1) and} \conf{\\}
   \co{(each x in s | e2)}
 \end{tabular}\\\hline   
7&
\begin{tabular}{@{}l@{}}
  \co{each x in s}  \conf{\\}
  \co{| e1 or e2}
\end{tabular}
 &
 \begin{tabular}{@{}l@{}}
   \co{each x in \{x in s | not e1\}} \conf{\\}
   \co{| e2}
 \end{tabular}\\\hline
8&
\begin{tabular}{@{}l@{}}
  \co{each x in s}  \conf{\\}
  \co{| e1 implies e2}  
\end{tabular}
 &
 \begin{tabular}{@{}l@{}}
   \co{each x in \{x in s | e1\}} \conf{\\}
   \co{| e2}
 \end{tabular}\\\hline
\end{tabular}
}

  \label{tab-extract}
\end{table}

\myparag{Updates caused by message passing}
Recall that the parameters of a query are variables in the query whose
values
are bound before the query.
Updates that may affect the query result include not only updates to the
query parameters but also updates to the objects and collections reachable
from the parameter values.
The most basic updates are assignments to query parameters,
\co{v\,:=\,exp}, where \co{v} is a query parameter.
Other updates are to objects and collections used in the query.
For objects, all updates can be expressed as field assignments,
\co{o.f\,:=\,exp}.
For collections, all updates can be expressed as initialization to empty
and element additions and removals, 
\co{s.add(x)} and \co{s.del(x)}.

For distributed algorithms, a distinct class of important updates are
caused by message passing.  Updates are caused in two ways:
\begin{enumerate}
\item Sending and receiving messages updates the sequences \co{sent}
  and \co{received}, respectively\notes{}.  Before incrementalization, code is generated, as
  described in Section~\ref{sec-compile}, to explicitly perform these
  updates.
  
\item Handling of messages by code in \co{receive} definitions updates
  variables that are parameters of the queries for computing
  synchronization conditions, or that are used to compute the values
  of these parameters.
\notes{}%
\end{enumerate}
Once these are established, updates can be determined using previously
studied analysis methods,
e.g.,~\cite{Liu+05OptOOP-OOPSLA,Gor+10Alias-DLS}.
\notes{}%

\myparag{Incremental computation}
Given expensive queries and updates to the query parameters, efficient
incremental computations can be derived for large classes of queries
and updates
based on the language constructs used in them
or by using a library of rules built on existing data
structures~\cite{PaiKoe82,Liu+05OptOOP-OOPSLA,Liu+06ImplCRBAC-PEPM,Liu+09Inv-GPCE}.

For aggregations converted from quantifications, algebraic properties
of the aggregation operators are exploited to efficiently handle
possible updates.
In particular, each resulting aggregate query result can be obtained
in \O{1} time and incrementally maintained in \O{1} time per update to
the sets maintained and affected plus the time for evaluating the
conditions in the aggregation once per update.
The total maintenance time at each element addition or deletion to a
query parameter is at least a linear factor smaller than computing the
query result from scratch.
Additionally, if aggregation operators \co{max} and \co{min} are used
and there are only element additions, %
the space overhead is \O{1}.
Note that if \co{max} and \co{min} are used naively when there are
element deletions, there may be an unnecessary overhead of \O{n} space
and \O{\log n} maintenance time per update from using more
sophisticated data structures to maintain the \co{max} or \co{min}
under element deletion~\cite{wil84range,wil02jcss,Cor+09}. %

Incremental computation improves time complexity only if the total
time of repeated expensive queries is larger than that of repeated
incremental maintenance.  This is generally true for incrementalizing
expensive synchronization conditions %
because
(1) expensive queries in the synchronization
conditions need to be evaluated repeatedly at each relevant update
to the message history, until the condition becomes true, and (2)
incremental maintenance at each such update is at least a linear
factor faster for single message updates and no slower generally than
computing from scratch.

To allow the most efficient incremental computation under all given
updates, our method transforms each top-level quantification as
follows:
\begin{itemize}

\item For non-nested quantifications, if the conditions contain no
  order comparisons or there are deletions from the sets or sequences
  whose elements are compared, the rules in Table~\ref{tab-quant} are
  used.  
  The space overhead is linear in the sizes of the sets
  maintained and being aggregated over.
 
\item For non-nested quantifications, if the conditions contain order
  comparisons and there are only additions to the sets or sequences
  whose elements are compared, the rules in Table~\ref{tab-extract}
  are used to extract single quantified order comparisons, and then
  the rules in Table~\ref{tab-cmp} are used to convert the extracted
  quantifications.  In this case, the space overhead is reduced to
  constant.

\item For nested quantifications with one level of nesting, the rules
  in Table~\ref{tab-extract} are used to extract directly nested
  quantifications, and then the rules in Table~\ref{tab-nest} are
  used.
  If the resulting incremental maintenance has constant-time overhead
  maintaining a linear-space structure, we are done.
  If it is linear-time overhead maintaining a quadratic-space
  structure, and if the conditions contain order comparisons, then the
  rules in Table~\ref{tab-extract} are used to extract single
  quantified order comparisons, and then the rules in
  Table~\ref{tab-cmp} are used.  This can reduce the overhead to
  logarithmic time and linear space.

\item In general, multiple ways of conversion may be possible, besides
  small constant-factor differences between rules~2 and 3 in
  Table~\ref{tab-quant} and rules~4 and 5 in Table~\ref{tab-nest}.  In
  particular, for nested quantifications with two or more
  alternations, one must choose which two alternating quantifiers to
  transform first, using rule~2 or 3 in Table~\ref{tab-nest}.
  We have not encountered such queries and have not studied this
  aspect further.  Our general method is to transform in all ways
  possible, obtain the time and space complexities for each result,
  and choose one with the best time and then space.  Complexities are
  calculated using the cost model of the set operations given in
  Section~\ref{sec-lang}.  The number of possible ways is exponential
  in the worst case in the size of the query, but the query size is
  usually a small constant.
  \notes{}

\end{itemize}
Table~\ref{tab-inc} summarizes well-known incremental computation
methods for these aggregate queries.  The methods are expressed as
incrementalization rules: if a query in the program matches the query
form in the table, and each update to a parameter of the query in the
program matches an update form in the table, then transform the query
into the corresponding replacement and insert at each update the
corresponding maintenance; fresh variables are introduced for each
different query to hold the query results or auxiliary data
structures.
In the third rule, data structure \co{ds} stores the argument set \co{s}
of \co{max} and supports priority queue operations.
\begin{table}[htbp]
\conf{
  \small
}
  \centering
  \caption{Incrementalization rules for \co{count} and for \co{max}.}
{
\begin{tabular}{@{~}l@{~}}
\begin{tabular}{l@{~~\hfill}|l@{\hfill}|l@{\hspace{6ex}}}\hline
\hspace{1.5ex}Query  &\hspace{10ex}Replacement\hspace{12ex} &Cost\\\hline
\co{count s}    &\co{number}  & \O{1}\\\hline\hline
~~Updates     &\hspace{6ex}Inserted Maintenance &Cost\\\hline
\co{s := \{\}}  &\co{number := 0} & \O{1}\\\hline
\co{s.add(x)}  &\co{if x not in s:~number +:= 1}  & \O{1}\\\hline
\co{s.del(x)}  &\co{if x in s:~number -:= 1}  & \O{1}\\\hline
\end{tabular}
\\\\
\begin{tabular}{l@{~~\hfill}|l@{\hfill}|l@{\hspace{6ex}}}\hline
\hspace{1.5ex}Query  &\hspace{10ex}Replacement\hspace{12ex} &Cost\\\hline
\co{max s}     &\co{maximum}    & \O{1}\\\hline\hline
~~Updates     &\hspace{6ex}Inserted Maintenance &Cost\\\hline
\co{s := \{x\}} &\co{maximum := x}  & \O{1}\\\hline
\co{s.add(x)}   &\co{if x > maximum:~maximum := x}  & \O{1}\\\hline
\end{tabular}
\\\\
\begin{tabular}{l@{~~\hfill}|l@{\hfill}|l}\hline
\hspace{1.5ex}Query  &\hspace{10ex}Replacement\hspace{12ex} &Cost\\\hline
\co{max s}     &\co{ds.max()}  & \O{1}\\\hline\hline
~~Updates     &\hspace{6ex}Inserted Maintenance &Cost\\\hline
\co{s := \{\}}  &\co{ds := new DS()}  & \O{1}\\\hline
\co{s := \{x\}} &\co{ds := new DS(); ds.add(x)}  & \O{1}\\\hline
\co{s.add(x)}  &\co{if x not in s:~ds.add(x)}  & \O{\log |\mbox{\co{s}}|}\\\hline
\co{s.del(x)}  &\co{if x in s:~ds.del(x)}  & \O{\log |\mbox{\co{s}}|}\\\hline
\end{tabular}  
\end{tabular}
}
\conf{
  The rule for \co{min} is similar to the rule for \co{max}.
}
  \label{tab-inc}
\end{table}
\notes{}

The overall incrementalization
algorithm~\cite{PaiKoe82,Liu+05OptOOP-OOPSLA,Liu+06ImplCRBAC-PEPM}
introduces new variables to store the results of expensive queries and
subqueries, as well as appropriate additional values, forming a set of
invariants, transforms the queries and subqueries to use the stored
query results and additional values, and transforms updates to query
parameters to also do incremental maintenance of the stored query
results and additional values.
\notes{}%

In particular, if queries are nested, inner queries are transformed
before outer queries.  Note that a comprehension such as \co{\{x in s |
  bexp\}} is incrementalized with respect to changes to parameters of
Boolean expression \co{bexp} as well as addition and removal of
elements of \co{s}; if \co{bexp} contains nested subqueries, then
after the subqueries are transformed, incremental maintenance of their
query results become additional updates to the enclosing query.  

At the end, variables and computations that are dead in the
transformed program are eliminated.
In particular, sequences \co{received} and \co{sent} will be eliminated
when appropriate, because queries using them have been compiled into
message handlers that only store and maintain values needed for
incremental evaluation of the synchronization conditions.

\myparag{Example}
In the program in Figure~\ref{fig-lam-orig}, three quantifications are
used in the synchronization condition in the \co{await} statement, and two
of them are nested.  The condition is copied below, except that
\co{('ack',t2,=p2) in received} is used in place of
\co{received('ack',t2,=p2)}.
\begin{code}
    each ('request', t2, p2) in q | 
      (t2,p2) != (t,self) implies (t,self) < (t2,p2)
    and each p2 in s | 
      some ('ack', t2, =p2) in received | t2 > t
\end{code}

Converting quantifications into aggregations as described using
Tables~\ref{tab-quant} through \ref{tab-extract} proceeds as follows.
In the first conjunct, the universal quantification is converted using
rule~2 or 3 in Table~\ref{tab-quant}, because it contains an order
comparison with elements of \co{q} and there are element deletions from
\co{q}; rule~3 is used here because it is slightly simpler after the
negated condition is simplified.
In the second conjunct, the nested quantification is converted using
rule~2 in Table~\ref{tab-nest}.
The resulting expression is:
\begin{code}
    count \{('request', t2, p2) in q |
            (t,self) > (t2,p2)\}  =  0
    and
    count \{p2: p2 in s, ('ack', t2, p2) in received |
            t2 > t\}  =  count s
\end{code}

Updates to parameters of the first conjunct are additions and removals
of requests to and from \co{q}, and also assignment to \co{t}.\notes{}
Updates to parameters of the second conjunct are additions of \co{ack}
messages to \co{received}, and assignment to \co{t},
after the initial assignment to \co{s}.

Incremental
computation~\cite{PaiKoe82,Liu+05OptOOP-OOPSLA,Liu+06ImplCRBAC-PEPM,Liu+09Inv-GPCE}
introduces variables to store the values of all three aggregations in
the converted query, transforms the aggregations to use the introduced
variables, and incrementally maintains the stored values at each of
the updates, as follows, yielding Figure~\ref{fig-lam-inc}.
\begin{itemize}

\item For the first conjunct, store the set value and the \co{count}
  value in two variables, say \co{earlier} and \co{number1},
  respectively,
  so first conjunct becomes \co{number1 = 0};
  when \co{t} is assigned a new value, let \co{earlier} be \co{q}
  and let \co{number1} be its size, taking \O{|\mbox{\co{earlier}}|}
  time, amortized to \O{1} time when each request in \co{earlier} is
  served;
  when a request is added to \co{q}, if \co{t} is defined and
  \co{(t,self) > (t2,p2)} holds, add the request to \co{earlier} and
  increment \co{number1} by 1, taking \O{1} time;
  similarly for deletion from \co{q}.
  A test of definedness, here \co{t != undefined}, is inserted
  for any variable that might not be %
  defined in the scope %
  of the maintenance code.

  Note that when \co{('request',t,self)} in particular is added to or
  removed from \co{q}, \co{earlier} and \co{number1} are not updated,
  because \co{(t,self) > (t,self)} is trivially false.

\item For the second conjunct, store the set value and the two
  \co{count} values in three variables, say \co{responded}, \co{number2},
  and \co{total}, respectively,
  so the conjunct becomes \co{number2 = total};
  when \co{s} is initialized in \co{setup}, assign \co{total} the size
  of \co{s}, taking \O{|\mbox{\co{s}}|} time, done only once for each
  process;
  when \co{t} is assigned a new value, let \co{responded} be \co{\{\}},
  and let \co{number2} be 0, 
  taking \O{1} time;
  when an \co{ack} message is added to \co{received}, if the associated
  conditions hold, increment \co{number2} by 1, taking \O{1} time.
  \notes{}
  A test of definedness of \co{t} is omitted in the maintenance for
  receiving \co{ack} messages, because \co{t} is always defined there;
  this small %
  optimization is %
  incorporated in an incrementalization rule, but it could be done
  with a
  data-flow analysis that covers distributed data flows.
\end{itemize}
Note that incrementalization uses basic properties about primitives
and libraries.  These properties are incorporated in
incrementalization rules.  For the running example, the property used
is that a call to \co{logical\_time()} returns a timestamp larger than
all
existing timestamp values, %
and thus at the assignment to \co{t} in method \co{mutex}, we have that
\co{earlier} is \co{q}
and \co{responded} is \co{\{\}}.
So, an incrementalization rule for maintaining \co{earlier}
specifies that at update \co{t := logical\_time()}, the maintenance is
\co{earlier := q}; similarly for maintaining \co{responded}.
These simplifications could be facilitated
with %
data-flow analyses that determine variables holding logical times and
sets holding certain element types.
Incrementalization rules can use any program analysis results as
conditions~\cite{Liu+09Inv-GPCE}.

Figure~\ref{fig-lam-inc} shows the optimized program after
incrementalization of the synchronization condition on lines 10-11 in
Figure~\ref{fig-lam-orig}.  All lines with comments are new except
that the synchronization condition in the \co{await} statement is
simplified.
The synchronization condition now takes \O{1} time, compared with
\O{|\mbox{\co{s}}|^2} if computed from scratch.  The trade-off is the
amortized \O{1} time overhead at updates to \co{t} and \co{q} and on
receiving of \co{ack} messages.
Using based representation for sets~\cite{Pai89,Cai+91,Goy00thesis},
maintaining \co{earlier} and \co{responded} can each be done using one
bit for each process.

Note that the sequence \co{received} used in the synchronization
condition in Figure~\ref{fig-lam-orig} is no longer used after
incrementalization.  All values needed for evaluating the
synchronization condition are stored in new variables introduced:
\co{earlier}, \co{number1}, \co{responded}, \co{number2}, and \co{total},
a drastic space improvement from unbounded for \co{received} to linear
in the number of processes.
\begin{figure}[htbp]
\begin{smallcode}
\tran{} 1 class P extends process:
\tran{} 2   def setup(s):
\tran{} 3     self.s := s
\tran{} 4     self.total := count s          # total num of other processes
\tran{} 5     self.q := \{\}
\tran{}
\tran{} 6   def mutex(task):
\tran{} 7     -- request
\tran{} 8     self.t := logical_time()
\tran{} 9     self.earlier := q              # set of pending earlier requests
\tran{}10     self.number1 := count earlier  # num of pending earlier requests
\tran{}11     self.responded := \{\}           # set of responded processes
\tran{}12     self.number2 := 0              # num of responded processes
\tran{}13     send ('request', t, self) to s
\tran{}14     q.add(('request', t, self))
\tran{}15     await number1 = 0 
\tran{}             and number2 = total      # use maintained results
\tran{}16     task()
\tran{}17     -- release
\tran{}18     q.del(('request', t, self))
\tran{}19     send ('release', logical_time(), self) to s
\tran{}
\tran{}20   receive ('request', t2, p2):
\tran{}21     if t != undefined:             # if t is defined
\tran{}22       if (t,self) > (t2,p2):       # comparison in conjunct 1
\tran{}23         if ('request',t2,p2) not in earlier:   # if not in earlier
\tran{}24           earlier.add(('request', t2, p2))     # add to earlier
\tran{}25           number1 +:= 1                        # increment number1
\tran{}26     q.add(('request', t2, p2))
\tran{}27     send ('ack', logical_time(), self) to p2
\tran{}
\tran{}28   receive ('ack', t2, p2):         # new message handler
\tran{}29     if t2 > t:                     # comparison in conjunct 2
\tran{}30       if p2 in s:                  # membership in conjunct 2
\tran{}31         if p2 not in responded:    # if not responded already
\tran{}32           responded.add(p2)        # add to responded
\tran{}33           number2 +:= 1            # increment number2
\tran{}
\tran{}34   receive ('release', _, p2):
\tran{}35     for ('request', t2, =p2) in q:
\tran{}36       if t != undefined:           # if t is defined
\tran{}37         if (t,self) > (t2,p2):     # comparison in conjunct 1
\tran{}38           if ('request',t2,p2) in earlier:     # if in earlier
\tran{}39             earlier.del(('request', t2, p2))   # delete from earlier
\tran{}40             number1 -:= 1                      # decrement number1
\tran{}41       q.del(('request', t2, p2))
\end{smallcode}

  \caption{Optimized program after incrementalization.
    Definitions of \co{run} and \co{main} are as in Figure~\ref{fig-lam-orig}.}
  \label{fig-lam-inc}
\end{figure}

\myparag{Example with naive use of aggregation operator min}
Note that the resulting program in Figure~\ref{fig-lam-inc} does not
need to use a queue at all, even though a queue is used in the
original description in Figure~\ref{fig-lam-paper}; the variable \co{q}
is simply a set, and thus element addition and removal takes \O{1} time.

We show that if \co{min} is used naively, a more sophisticated data
structure~\cite{wil84range,wil02jcss,Cor+09} %
supporting priority queue is needed, incurring an \O{\log n} time
update instead of the \O{1} time in Figure~\ref{fig-lam-inc}.
Additionally, for a query using \co{min} to be correct, special care
must be taken to deal with the case when the argument to \co{min} is
empty, because then \co{min} is undefined.

Consider the first conjunct in the synchronization condition in the
\co{await} statement in Figure~\ref{fig-lam-orig}, copied below:
\begin{code}
\tran{}  each ('request', t2, p2) in q | 
\tran{}    (t2,p2) != (t,self) implies (t,self) < (t2,p2)
\end{code}
One might have written the following instead, because it seems natural,
especially if universal quantification is not supported:
\begin{code}
\tran{}  (t,self) < min \{(t2,p2): ('request', t2, p2) in q
\tran{}                         | (t2,p2) != (t,self)\}
\end{code}
However, that is incorrect, because the argument of \co{min} may be
empty, in which case \co{min} is undefined.

Instead of resorting to commonly used special values, such as
\co{maxint}, which is ad hoc and error prone in general, the empty case
can be added as the first disjunct of a disjunction:\pagebreak
\begin{code}
\tran{}  \{(t2,p2): ('request', t2, p2) in q
\tran{}          | (t2,p2) != (t,self)\} = \{\}
\tran{}  or
\tran{}  (t,self) < min \{(t2,p2): ('request', t2, p2) in q
\tran{}                         | (t2,p2) != (t,self)\}
\end{code}
In fact, the original universal quantification in the first conjunct
in the \co{await} statement can be converted exactly to this
disjunction by using rule~8 in Table~\ref{tab-extract} and then
rule~13 in Table~\ref{tab-cmp}.  Our method does not consider this
conversion because it leads to a worse resulting program.

Figure~\ref{fig-lam-inc-min} shows the resulting program after
incrementalization of the synchronization condition that uses the
disjunction above, where \co{ds} stores the argument set of \co{min} and
supports priority queue operations.  All commented lines are new
compared to Figure~\ref{fig-lam-orig} except that the synchronization
condition in the \co{await} statement is simplified.
The program appears shorter than Figure~\ref{fig-lam-inc} because the
long complex code for maintaining the data structure \co{ds} is not
included; it is in fact similar to Figure~\ref{fig-lam-inc} except
that \co{ds} is used and maintained instead of \co{earlier} and
\co{number1}.

The program in Figure~\ref{fig-lam-inc-min} is still a drastic
improvement over the original program in Figure~\ref{fig-lam-orig},
with the synchronization condition reduced to \O{1} time and with
\co{received} removed, just as in Figure~\ref{fig-lam-inc}.
The difference is that maintaining \co{ds} for incrementalizing \co{min}
under element addition to and deletion from \co{q} takes \O{\log
  |\mbox{\co{s}}|} time, as opposed to \O{1} time for maintaining
\co{earlier} and \co{number1} in Figure~\ref{fig-lam-inc}.
\begin{figure}[htbp]
\begin{smallcode}
\tran{} 1 class P extends process:
\tran{} 2   def setup(s):
\tran{} 3     self.s := s
\tran{} 4     self.total := count s          # total num of other processes
\tran{} 5     self.q := \{\}
\tran{} 6     self.ds := new DS()            # data structure for maintaining
\tran{}                                      #  requests by other processes
\tran{} 7   def mutex(task):
\tran{} 8     -- request
\tran{} 9     self.t := logical_time()
\tran{}10     self.responded := \{\}           # set of responded processes
\tran{}11     self.number := 0               # num of responded processes
\tran{}12     send ('request', t, self) to s
\tran{}13     q.add(('request', t, self))
\tran{}14     await (ds.is_empty() or (t,self) < ds.min())
\tran{}             and number = total       # use maintained results
\tran{}15     task()
\tran{}16     -- release
\tran{}17     q.del(('request', t, self))  
\tran{}18     send ('release', logical_time(), self) to s
\tran{}
\tran{}19   receive ('request', t2, p2):
\tran{}20     ds.add((t2,p2))                # add to data structure
\tran{}21     q.add(('request', t2, p2))  
\tran{}22     send ('ack', logical_time(), self) to p2
\tran{}
\tran{}23   receive ('ack', t2, p2):         # new message handler
\tran{}24     if t2 > t:                     # comparison in conjunct 2
\tran{}25       if p2 in s:                  # membership in conjunct 2
\tran{}26         if p2 not in responded:    # if not responded already
\tran{}27           responded.add(p2)        # add to responded
\tran{}28           number +:= 1             # increment number
\tran{}
\tran{}29   receive ('release', _, p2):
\tran{}30     for ('request', t2, =p2) in q:
\tran{}31       ds.del((t2,p2))              # delete from data structure
\tran{}32       q.del(('request', t2, p2))  
\end{smallcode}
\caption{Optimized program with use of \co{min} after incrementalization.
  Definitions of \co{run} and \co{main} are as in Figure~\ref{fig-lam-orig}.}
  \label{fig-lam-inc-min}
\end{figure}

\myparag{Simplifications to the original algorithm}
Consider the original algorithm in Figure~\ref{fig-lam-orig}.  Note
that incrementalization determined that there is no need for a process
to update auxiliary values for its own request, in both
Figures~\ref{fig-lam-inc} and~\ref{fig-lam-inc-min}.  Based on this,
we discovered, manually, that updates to \co{q} for a process's own
request do not affect the two uses of \co{q}, on lines 9 and 35, in
Figure~\ref{fig-lam-inc} and the only use of \co{q}, on line 30, in
Figure~\ref{fig-lam-inc-min}.  So we can remove them in
Figures~\ref{fig-lam-inc} and~\ref{fig-lam-inc-min}.  In addition, we
can remove them on lines 9 and 14 in Figure~\ref{fig-lam-orig} and
remove the test \co{(t2,p2) != (t,self)}, which becomes always true, in
the synchronization condition, yielding a simplified original
algorithm.

Furthermore, note that the remaining updates to \co{q} in
Figure~\ref{fig-lam-orig} merely maintain pending requests by others,
so we can remove lines 4, 17, 20, 21, and the entire \co{receive}
definition for \co{release} messages, by using, for the first conjunct
in the \co{await} statement,
\begin{code}
    each received('request', t2, p2) |
      not (some received('release', t3, =p2) | t3 > t2)
      implies (t,self) < (t2,p2)
\end{code}

Figure~\ref{fig-lam-simp} shows the resulting simplified algorithm.
Incrementalizing this program yields essentially the same programs as in
Figures~\ref{fig-lam-inc} and~\ref{fig-lam-inc-min}, 
except that it needs to use the property that when a message is added
to \co{received}, 
all messages from the same process in \co{received} have a smaller timestamp.
This property follows from the use of logical clock and FIFO channels.
The incrementalization rules for maintaining the result of the new
condition incorporate this property in a similar way as described for
Figure~\ref{fig-lam-inc}, except it could be facilitated with also a
data-flow analysis that determines the component of a received message
holding the sender of the message.

\begin{figure}[htbp]
\begin{smallcode}
\tran{} 1 class P extends process:
\tran{} 2   def setup(s):
\tran{} 3     self.s := s
\tran{}
\tran{} 4   def mutex(task):
\tran{} 5     -- request
\tran{} 6     self.t := logical_time()
\tran{} 7     send ('request', t, self) to s
\tran{} 8     await each received('request', t2, p2) |
\tran{}               not (some received('release', t3, =p2) | t3 > t2)
\tran{}               implies (t,self) < (t2,p2)
\tran{} 9           and each p2 in s |
\tran{}               some received('ack', t2, =p2) | t2 > t
\tran{}10     task()
\tran{}11     -- release
\tran{}12     send ('release', logical_time(), self) to s
\tran{}
\tran{}13   receive ('request', _, p2):
\tran{}14     send ('ack', logical_time(), self) to p2
\end{smallcode}
  \caption{Simplified algorithm.
    Definitions of \co{run} and \co{main} are as in
    Figure~\ref{fig-lam-orig}.}
  \label{fig-lam-simp}
\end{figure}

\msg{}

\msg{}

\section{Implementation and experiments}
\label{sec-expe}

We have developed a prototype implementation of the compiler and
optimizations for DistAlgo and evaluated it in implementing a set of
well-known distributed algorithms, as described
previously~\cite{Liu+12DistPL-OOPSLA}.
We have also used DistAlgo in teaching distributed algorithms and
distributed systems, and students used the language and system in
programming assignments and course projects.
We summarize results from the former and describe experience with the
latter, after an overview and update about the implementation.

Our DistAlgo implementation takes DistAlgo programs written in
extended Python, applies analyses and optimizations, especially to the
high-level queries, and generates executable Python code.  It
optionally interfaces with an incrementalizer to apply
incrementalization before generating code.
Applying incrementalization uses the methods and implementation from
previous
work:
a library of incrementalization rules was developed, manually but
mostly following a systematic
method~\cite{Liu+05OptOOP-OOPSLA,Liu+06ImplCRBAC-PEPM}, and applied
automatically using InvTS~\cite{Liu+09Inv-GPCE,Gor+10Alias-DLS}.
A set of heuristics %
are currently used to select the best program generated from
incrementalizing differently converted aggregations.

A more extensive implementation of DistAlgo than the first
prototype~\cite{Liu+12DistPL-OOPSLA} has been released and is being
gradually improved~\cite{distalgo}.
Improved methods and implementation for incrementalization are also
being developed~\cite{Liu+16IncOQ-PPDP},
to replace manually written
incrementalization rules, and to better select the best transformed
programs.

\myparag{Evaluation in implementing distributed algorithms}
We have used DistAlgo to implement a variety of well-known distributed
algorithms, including twelve different algorithms for distributed
mutual exclusion\notes{}, %
leader election\notes{}, %
and atomic commit\notes{}, %
as well as Paxos\notes{}, %
Byzantine Paxos\notes{}, %
and %
multi-Paxos\notes{}, %
as summarized previously~\cite{Liu+12DistPL-OOPSLA}; results of
evaluation using these programs are as follows:
\begin{itemize}
\item DistAlgo programs are consistently small, ranging from 22 to 160
  lines,
  and are much smaller than specifications or programs written in
  other languages, mostly 1/2 to 1/5 of the size;
  also we were able to find only a few of these algorithms written in
  other languages.
  Our own best effort to write Lamport's distributed mutual exclusion
  in programming languages resulted in 272 lines in C, 216 lines in
  Java, 122 lines in Python, and 99 lines in Erlang, compared with 32
  lines in DistAlgo.

\item Compilation times without incrementalization are all under 0.05
  seconds on an Intel Core-i7 2600K CPU with 16GB of memory; and
  incrementalization times are all under 30 seconds.
  Generated code size ranges from 1395 to 1606 lines of Python,
  including 1300 lines of fixed library code.

\item Execution time and space confirm the analyzed asymptotic time
  and space complexities.  For example, for Lamport's distributed
  mutual exclusion, total CPU time is linear in the number of
  processes for the incrementalized program, but superlinear for the
  original program; for a fixed number of processes, the memory usage
  is constant for the incremental program, but grows linearly with the
  number of requests for the original program.

\item Compared with running times of our best, manually written
  programs in programming languages, all running on a single machine,
  our generated DistAlgo takes about twice as long as our Python
  version, which takes about twice as long as our Java version, which
  takes about twice as long as our C version, which takes about four
  times as long as our Erlang version.
  
\end{itemize}
Python is well known to be slow compared Java and C, and we have not
focused on optimizing constant factors.  Erlang is significantly
faster than C and the rest because of its use of light-weight threads
to implement processes that is facilitated by its being a functional
language.  However, among all our programs for Lamport's distributed
mutual exclusion, Erlang is the only one besides un-incrementalized
DistAlgo whose memory usage for a fixed number of processes grows
linearly with the number of requests.

Programming distributed algorithms at a high level has also allowed us
to discover several improvements to correctness and efficiency aspects
of some of the algorithms~\cite{Liu+12DistSpec-SSS}.
For example, in the\notes{} pseudocode for
multi-Paxos~\cite{vra15paxos}, in process \co{Commander}, waiting for
\co{p2b} messages containing ballot \co{b} from a majority of
\co{acceptors} is expressed by starting with a \co{waitfor} set
initialized to \co{acceptors} and then, in a \co{for ever} loop,
repeatedly updating \co{waitfor} and testing \co{|waitfor| <
  |acceptors|/2} as each \co{p2b} message containing ballot \co{b}
arrives.  The test is incorrect if implemented directly in commonly
used languages such as Java, and even Python until Python 3, because
\co{/} is integer division, which discards any fractional part; for
example, test \co{1\,<\,3/2} becomes \co{false} but should be \co{true}.
In DistAlgo, the entire code can simply be written as
\begin{code}
    await count \{a: received ('p2b',=b) from a\} > \conf{\newline}\conf{\mbox{          }}(count acceptors)/2
\end{code}
using the standard majority test, and it is correct whether \co{/} is
for integer or float.

\myparag{Experience in teaching distributed algorithms}
DistAlgo has also helped us tremendously in teaching distributed
algorithms, because it makes complex algorithms completely clear,
precise, and directly executable.
Students learn DistAlgo quickly through even a small programming
assignment, despite that most did not know Python before, thanks to
the power and clarity of Python.

In particular, students in distributed systems courses have used
DistAlgo in dozens of course projects, implementing the core of
network protocols and distributed graph algorithms~\cite{Lynch96};
distributed coordination services Chubby~\cite{burrows06chubby} and
Zookeeper~\cite{hunt2010zookeeper}; distributed hash tables
Kademlia~\cite{maymounkov2002kademlia}, Chord~\cite{stoica2003chord},
Pastry~\cite{rowstron2001pastry}, Tapestry~\cite{zhao2004tapestry},
and Dynamo~\cite{decandia2007dynamo}; distributed file systems
GFS~\cite{ghemawat2003google} and HDFS~\cite{shvachko2010hadoop};
distributed databases Bigtable~\cite{chang2008bigtable},
Cassandra~\cite{lakshman2010cassandra}, and
Megastore~\cite{baker11megastore}; distributed processing platform
MapReduce~\cite{dean2008mapreduce}; and others.

All distributed programming features were used extensively in
students' programs---easy process creation and setup and sending of
messages, high-level control flows with \co{receive} definitions as
well as \co{await} for synchronization, and declarative
configurations---with the exception of queries over message histories,
because students had been trained in many courses to handle events
imperatively; we have not evaluated incrementalization on students'
programs, because execution efficiency has not been a problem.
Overall, students' experience helps confirm that DistAlgo allows
complex distributed algorithms and services to be implemented much
more easily than commonly used languages such as C++ and Java.
We summarize two specific instances below.

In a graduate class in Fall 2012, most of the 28 students initially
planned to use Java or C++ for their course projects, because they
were familiar with those and wanted to strengthen their experience of
using them instead of using DistAlgo in implementing distributed
systems.  However, after doing one programming assignment using
DistAlgo, all those students switched to DistAlgo for their course
projects,
except for one student, %
who had extensive experience with C++, including several years of
internship at Microsoft Research programming distributed systems.
\begin{itemize}
\item This student wrote about 3000 lines of C++, compared to about
  300 lines of DistAlgo written by several other students who chose
  the same project of implementing multi-Paxos and several
  optimizations.
  Furthermore, his C++ program was incomplete, lacking some
  optimizations that other students' DistAgo programs included.

\item The student did a re-implementation in DistAlgo quickly after
  the course\footnote{The student wanted to do research on DistAlgo
    and so was asked to re-implement his project in DistAlgo.},
  confirming that it took about 300 lines.
  His biggest surprise was that his C++ program was %
  an order of magnitude slower than his DistAlgo program.
  After several weeks of debugging, he found that it was due to an
  improper use of some C++ library %
  function.
\end{itemize}
The main contrast that the student concluded was the huge advantage of
DistAlgo over C++ in ease of programming and program understanding,
not to mention the unexpected performance advantage.

In a graduate class %
in Fall 2014, each team of two students first implemented a
fault-tolerant banking service in two languages: DistAlgo and another
language of their choice other than Python.  We excluded Python as the
other language, because implementing the same service in such closely
related languages would be less educational.  The service uses chain
replication \cite{vanrenesse04chain} to tolerate crash failures.  The
service offers only a few simple banking operations (get balance,
deposit, withdrawal, intra-bank transfer, inter-bank transfer), so
most of the code is devoted to distributed systems aspects.  The
numbers of teams that chose various other languages are: Java 15, C++
3, Go 3, Erlang 2, Node.js 2, Elixir (a variant of Erlang) 1,
JavaScript~1.
\begin{itemize}
\item

  In the last assignment, teams implemented an extension to the
  banking service in one language of their choice.  59\% of the teams
  chose DistAlgo for this, even though most students (about 80\%) did
  not know Python, and none knew DistAlgo, at the beginning of the
  class.  In other words, a majority of students decided that
  implementation of this type of system is
  better\notes{} in DistAlgo, even
  compared to languages with which they had more experience and that
  are more widely used.

\item

  We asked each team to compare their experiences with the two
  languages.  Teams consistently reported that development in DistAlgo
  was faster and easier than development in the other language (even
  though most students did not know Python before the project), and
  that the DistAlgo code was significantly shorter.  It is no surprise
  that Java and C++ require more code, even when students used
  existing networking libraries, which they were encouraged to do.
  Comparison with Erlang and Go is more interesting, because they are
  high-level languages designed to support distributed programming.
  For the teams that chose Erlang, the average DistAlgo and Erlang
  code sizes, measured as non-empty non-comment line of code, are 586
  and 1303, respectively.  For the teams that chose Go, the average
  DistAlgo and Go code sizes are 465 and 1695, respectively.

\end{itemize}

\section{Related work and conclusion}
\label{sec-related}

A wide spectrum of languages and notations have been used to describe
distributed algorithms,
e.g.,~\cite{Ray88book,Lynch96,Lam02book,Garg02,AttWel04,KshSin08,Lam09pluscal,Tel00,Ray10async,Ray13dist}\notes{}.
At one end, pseudocode with English is used, e.g.,~\cite{KshSin08},
which gives a high-level flow of the algorithms, but lacks the details
and precision needed for a complete understanding.  At the other end,
state machine based specification languages are used, e.g., I/O
automata~\cite{Lynch96,Kaynar10TIOA}, which is completely precise, but
uses low-level control flows that make it harder to write and
understand the algorithms.  There are also many notations in between
these extremes, some being much more precise or completely precise
while also giving a high-level control flow, e.g., Raynal's
pseudocode~\cite{Ray88book,Ray10async,Ray13dist} and Lamport's
PlusCal~\cite{Lam09pluscal}.  However, all of these languages and
notations lack concepts and mechanisms for building real distributed
applications, and most of the languages are not executable\notes{}. %

Many programming languages support programming of distributed
algorithms and applications.
Most support distributed programming through messaging libraries,
ranging from relatively simple socket libraries to complex libraries
such as MPI~\cite{MPI_Forum}.  Many support Remote Procedure Call
(RPC) or Remote Method Invocation (RMI), which allows
a process to call a subroutine in another process without the
programmer coding the details for this.  
Many also support asynchronous method invocation (AMI), which allows
the caller to not block and get the reply later.
Some programming languages, such as Erlang~\cite{larson09erlang,erlang},
which has an actor-like model~\cite{agha1986actors}, have support for
message passing and process management built into the language.
There are also other well-studied languages for distributed
programming, e.g., Argus~\cite{liskov88argus},
Lynx~\cite{scott91lynx},
SR~\cite{andrews1993sr},
Concert/C~\cite{auer94concert}, and
Emerald~\cite{black07emerald}. %
These languages all lack constructs for expressing control flows and
complex synchronization conditions at a much higher level; such
high-level constructs are extremely difficult to implement
efficiently.
DistAlgo's construct for declaratively and precisely specifying yield
points for handling received messages is a new feature that we have
not seen in other languages.
So is DistAlgo's support of history variables in high-level
synchronization conditions in non-deterministic \co{await} with
timeout in a programming language.
Our simple combination of synchronous \co{await} and asynchronous
\co{receive} allows distributed algorithms to be expressed easily and
clearly.

There has been much work on producing executable implementations from
formal specifications, e.g., from process algebras~\cite{HanCleSmo04},
I/O automata~\cite{georgiou09automated}, Unity~\cite{Granicz03unity},
and Seuss~\cite{Kruger96seuss},
as well as from more recently proposed high-level languages for
distributed algorithms, e.g., Datalog-based languages
Meld~\cite{ashley-rollman-iclp09}, Overlog~\cite{alvaro2010declare},
and Bloom~\cite{bloom}, a Prolog-based language
DAHL~\cite{lopes2010applying}, and a logic-based language
EventML~\cite{Bic09,eventml12}.
An operational semantics was studied recently for a variant of Meld,
called Linear Meld, that allows updates to be encoded more
conveniently than Meld by using linear logic~\cite{cruz14LM}.
Compilation of DistAlgo to executable implementations is easy because
it is designed to be so and DistAlgo is given an operational
semantics.  High-level queries and quantifications used for
synchronization conditions
can be compiled into loops straightforwardly, but they may be
extremely inefficient.  None of these prior works study powerful
optimizations of quantifications.  Efficiency concern is a main reason
that similar high-level language constructs, whether for queries or
assertions, are rarely used, if supported at all, in commonly used
languages.

Incrementalization has been studied extensively,
e.g.,~\cite{RamRep93,Liu13book}, 
both for doing it systematically based on languages, and in applying it
in an ad hoc fashion to specific problems.
However, all systematic incrementalization methods based on languages
have been for centralized sequential programs, e.g., for
loops~\cite{Allen:Cocke:Kennedy:81,Liu+05Array-TOPLAS,GauRaj06}, set
languages~\cite{PaiKoe82,GupMumSub93,Liu+06ImplCRBAC-PEPM}, recursive
functions~\cite{Pugh:Teitelbaum:89,LiuSto03DynProg-HOSC,Acar06adaptive},
logic rules~\cite{SahaRam03,LiuSto09Rules-TOPLAS}, and object-oriented
languages~\cite{Nak01,Liu+05OptOOP-OOPSLA,RotLiu08OSQ-GPCE,Liu+16IncOQ-PPDP}.
This work is the first to extend incrementalization to distributed
programs to support high-level synchronization conditions.
This allows the large body of previous work on incrementalization,
especially on sets and sequences, to be used for optimizing
distributed programs.

Quantifications are the centerpiece of first-order logic, and are
dominantly used in writing synchronization conditions and assertions
in specifications, but
there are few results on generating efficient implementations of them.
In the database area, despite extensive work on efficient
implementation of high-level queries, efficient implementation of
quantification has only been studied in limited scope or for extremely
restricted query forms,
e.g.,~\cite{Badia+95query,Cla+97quant,Badia07question,Badia+08quant}.
In logic programming, handling of universal quantification
is\notes{} based on variants of brute-force Lloyd-Topor
transformations, e.g.,~\cite{Pet97quantified,Fio+11progtrans}; even
state-of-the-art logic programming systems, e.g.,~\cite{xsb16}, do not
support universal quantification.
Our method is the first general and systematic method for
incrementalizing arbitrary quantifications.  Although they are much
more challenging to optimize than set queries, our method combines a
set of general transformations to transform them into aggregations
that can be most efficiently incrementalized using the best previous
methods.

\msg{}

\notes{}%

To conclude, this article presents a powerful language and method for
programming and optimizing distributed algorithms.  There are many
directions for future work, from formal verification on the theoretical
side, to generating code in lower-level languages on the practical side,
with many additional analyses and optimizations in between.
In particular, a language with a high level of abstraction also
faciliates formal verification,
of not only the high-level programs, but also the generated efficient
implementations when they are generated through systematic
optimizations.
Besides developing systematic optimizations,
we have started to study formal verification of distributed
algorithms~\cite{Cha+16PaxosTLAPS-FM} and their implementations by
starting with their high-level, concise descriptions in DistAlgo.
\notes{}%

\appendix
\section*{APPENDIX}

\newcommand{\mysection}[1]{\subsection{#1}}
\newcommand{\appendixOnly}[1]{#1}
\newcommand{\standaloneOnly}[1]{}

\setcounter{topnumber}{2}
\setcounter{bottomnumber}{2}
\setcounter{totalnumber}{4}     %
\setcounter{dbltopnumber}{2}

\newcommand{\firstalt}{~~}
\newcommand{\alt}{~~}

\newcommand{\pfn}{\rightharpoonup}
\newcommand{\bijection}{\stackrel{1-1}{\rightarrow}}
\newcommand{\union}{\cup}
\newcommand{\intersect}{\cap}

\newcommand{\Set}[1]{{\rm Set}(#1)}
\newcommand{\ra}{\rightarrow}
\newcommand{\myiff}{\Leftrightarrow}

\newcommand{\Bool}{{\it Bool}}
\newcommand{\Int}{{\it Int}}
\newcommand{\Address}{\mathify{\it Address}}
\newcommand{\ProcessAddress}{{\it ProcessAddress}}
\newcommand{\NonProcessAddress}{{\it NonProcessAddress}}
\newcommand{\Val}{\mathify{\it Val}}
\newcommand{\Object}{{\it Object}}
\newcommand{\SetOfVal}{{\it SetOfVal}}
\newcommand{\SeqOfVal}{{\it SeqOfVal}}
\newcommand{\MsgQueue}{{\it MsgQueue}}
\newcommand{\ChannelStates}{{\it ChannelStates}}
\newcommand{\LocalHeap}{{\it LocalHeap}}
\newcommand{\Heap}{{\it Heap}}
\newcommand{\HeapType}{{\it HeapType}}
\newcommand{\State}{{\it State}}
\newcommand{\Program}{\mathify{\it Program}}
\newcommand{\Configuration}{\mathify{\it Configuration}}
\newcommand{\ProcessClass}{\mathify{\it ProcessClass}}
\newcommand{\Method}{\mathify{\it Method}}
\newcommand{\ReceiveDef}{\mathify{\it ReceiveDef}}
\newcommand{\ReceivePattern}{\mathify{\it ReceivePattern}}
\newcommand{\Pattern}{\mathify{\it Pattern}}
\newcommand{\InstanceVariable}{\mathify{\it InstanceVariable}}
\newcommand{\MethodName}{\mathify{\it MethodName}}
\newcommand{\Parameter}{\mathify{\it Parameter}}
\newcommand{\Expression}{\mathify{\it Expression}}
\newcommand{\Iterator}{\mathify{\it Iterator}}
\newcommand{\AnotherAwaitClause}{\mathify{\it AnotherAwaitClause}}
\newcommand{\Literal}{\mathify{\it Literal}}
\newcommand{\BooleanLiteral}{\mathify{\it BooleanLiteral}}
\newcommand{\IntegerLiteral}{\mathify{\it IntegerLiteral}}
\newcommand{\UnaryOp}{\mathify{\it UnaryOp}}
\newcommand{\BinaryOp}{\mathify{\it BinaryOp}}
\newcommand{\TuplePattern}{\mathify{\it TuplePattern}}
\newcommand{\PatternElement}{\mathify{\it PatternElement}}
\newcommand{\ChannelOrder}{\mathify{\it ChannelOrder}}
\newcommand{\ChannelReliability}{\mathify{\it ChannelReliability}}
\newcommand{\EC}{\mathify{\it C}}

\newcommand{\ClassName}{\mathify{\it ClassName}}
\newcommand{\Label}{\mathify{\it Label}}
\newcommand{\Field}{\mathify{\it Field}}
\newcommand{\Statement}{\mathify{\it Statement}}
\newcommand{\Tuple}{\mathify{\it Tuple}}

\newcommand{\commentMark}{/\,/}

\newcommand{\commentS}[1]{\mbox{\commentMark\ #1}}

\newcommand{\tuple}[1]{(#1)}
\newcommand{\ltuple}[1]{(#1}
\newcommand{\rtuple}[1]{#1)}
\newcommand{\seq}[1]{\langle#1\rangle}
\newcommand{\set}[1]{\{#1\}}

\newcommand{\new}{{\it new}}
\newcommand{\emptySet}{\set{}}
\newcommand{\emptyfn}{f_0}
\newcommand{\emptyseq}{\seq{}}
\newcommand{\IF}{\mbox{if }}
\newcommand{\dom}{{\it dom}}
\newcommand{\rest}{{\it rest}}
\newcommand{\length}{{\it length}}
\newcommand{\first}{{\it first}}
\newcommand{\iscopy}{{\it isCopy}}
\newcommand{\matchRcvDef}{{\it matchRcvDef}}
\newcommand{\rcvAtLabel}{{\it receiveAtLabel}}

\newcommand{\bottom}{\mathord{\perp}}
\newcommand{\extends}{{\it extends}}
\newcommand{\methodDef}{{\it methodDef}}
\newcommand{\addrs}{{\it addrs}}
\newcommand{\subst}{{\it subst}}
\newcommand{\Init}{{\it Init}}
\newcommand{\sci}{\hspace*{0.75em}}
\newcommand{\spce}{\hspace*{1.5em}}

\standaloneOnly{}

\appendixOnly{
\section{Semantics of DistAlgo}

We give an abstract syntax and operational semantics for a core language
for DistAlgo.  The operational semantics is a reduction semantics with
evaluation contexts \cite{wright94syntactic,serbanuta07rewriting}.  
}

\mysection{Abstract Syntax}
\label{sec:syntax}

The abstract syntax is defined in Figures \ref{fig:syntax1} and
\ref{fig:syntax2}.  
We use some syntactic sugar in sample code, e.g., we use infix notation for
some binary operators, such as {\tt and} and {\tt is}.

\newenvironment{ctabbing}
          {\begin{center}\begin{minipage}{\textwidth}\begin{tabbing}}
          {\end{tabbing}\end{minipage}\end{center}}

\begin{figure*}[htb]
\setstretch{0.9}
\hspace*{1em}
\begin{ctabbing}
\Program\ ::= \Configuration\ \ProcessClass*\ \Method\\

\ProcessClass\ ::= {\tt class} \ClassName\ {\tt extends} \ClassName: \Method* \ReceiveDef*\\
\\
\ReceiveDef\ ::= \=
 {\tt receive} \ReceivePattern+ {\tt at} \Label+ {\tt :} \Statement\\
\> {\tt receive} \ReceivePattern+ {\tt :} \Statement\\
\\
\ReceivePattern\ ::= \Pattern\ {\tt from} \InstanceVariable\\
\\
\Method\ ::= \=
 {\tt def} \MethodName{\tt (}\Parameter*{\tt )} \Statement\\
\> {\tt defun} \MethodName{\tt (}\Parameter*{\tt )} \Expression\\
\\
\Statement\ ::=\\
\firstalt \InstanceVariable\ {\tt :=} \Expression\\
\alt \InstanceVariable\ {\tt :=} {\tt new} \ClassName\\
\alt \InstanceVariable\ {\tt := \{} \Pattern\ {\tt :} \Iterator* {\tt |} \Expression\ {\tt \}}\\
\alt \Statement\ {\tt ;} \Statement\\
\alt {\tt if} \Expression{\tt :} \Statement\ {\tt else:} \Statement\\
\alt {\tt for} \Iterator{\tt :} \Statement\\
\alt {\tt while} \Expression{\tt :} \Statement\\
\alt \Expression{\tt .}\MethodName{\tt (}\Expression*{\tt )}\\
\alt  {\tt send} \Tuple\ {\tt to} \Expression\\
\alt \Label\ {\tt await} \Expression\ {\tt :} \Statement\ \AnotherAwaitClause*\\
\alt \Label\ {\tt await} \Expression\ {\tt :} \Statement\ \AnotherAwaitClause* {\tt timeout} \Expression\\
\alt {\tt skip}\\  %
\\
\Expression\ ::= \=
\Literal\\
\> \Parameter\\
\> \InstanceVariable\\
\> \Tuple\\
\> \Expression{\tt .}\MethodName{\tt (}\Expression*{\tt )}\\
\> \UnaryOp{\tt (}\Expression{\tt )}\\
\> \BinaryOp{\tt (}\Expression{\tt ,}\Expression{\tt )}\\
\> {\tt isinstance}{\tt (}\Expression,\ClassName{\tt )}\\
\> {\tt and}{\tt (}\Expression,\Expression{\tt )} \spce \= \commentMark\ conjunction (short-circuiting)\\
\> {\tt or}{\tt (}\Expression,\Expression{\tt )}        \> \commentMark\ disjunction (short-circuiting)\\
\> {\tt each} \Iterator\ {\tt |} \Expression\\
\> {\tt some} \Iterator\ {\tt |} \Expression\\
\\
\Tuple\ ::= {\tt (}\Expression*{\tt )}
\end{ctabbing}
\caption{Abstract syntax, Part 1.}
\label{fig:syntax1}
\end{figure*}

\begin{figure*}[htb]
\setstretch{0.9}
\begin{ctabbing}
\UnaryOp\ ::= \=
   {\tt not}    \hspace*{3em} \= \commentMark\ Boolean negation\\
\> {\tt isTuple}           \> \commentMark\ test whether a value is a tuple\\
\> {\tt len}               \> \commentMark\ length of a tuple\\

\BinaryOp\ ::= \=
   {\tt is}           \> \commentMark\ identity-based equality\\
\> {\tt plus}         \> \commentMark\ sum\\
\> {\tt select}       \> \commentMark\ {\tt select}($t$,$i$) returns the $i$'th component of tuple $t$\\
\\
\Pattern\ ::= \=
   \InstanceVariable\\
\> \TuplePattern\\
\\
\TuplePattern\ ::= {\tt (}\PatternElement*{\tt )}\\
\\
\PatternElement\ ::= \=
   \Literal\\
\> \InstanceVariable\\
\> {\tt =}\InstanceVariable\\
\\
\Iterator\ ::= \Pattern\ {\tt in} \Expression\\
\\
\AnotherAwaitClause\ ::= {\tt or} \Expression\ {\tt :} \Statement\\
\\
\Configuration\ ::= {\tt configuration} \ChannelOrder\ \ChannelReliability\ ...\\
\ChannelOrder\ ::= \=
   {\tt fifo}\\
\> {\tt unordered}\\
\ChannelReliability\ ::= \=
   {\tt reliable}\\
\> {\tt unreliable}\\
\\
\ClassName\ ::= %
...\\
\MethodName\ ::= ...\\
\Parameter\ ::= %
...\\
\InstanceVariable\ ::= \Expression.\Field\\
\Field\ ::= ...\\
\Label\ ::= ...\\
\Literal\ ::= \=
  \BooleanLiteral\\
\> \IntegerLiteral\\
\> ...\\
\BooleanLiteral\ ::= \=
   {\tt true}\\
\> {\tt false}\\
\IntegerLiteral\ ::= ...
\end{ctabbing}
  \caption{Abstract syntax, Part 2. Ellipses (``...'') are for
    common syntactic categories whose details are unimportant.}
\label{fig:syntax2}
\end{figure*}

\myparag{Notation}

\begin{itemize}
\item A symbol in the grammar is a terminal symbol if it starts with a
  lower-case letter.
\item A symbol in the grammar is a non-terminal symbol if it starts with an
  upper-case letter.
\item In each production, alternatives are separated by a linebreak.
\item {\tt *} after a non-terminal means ``0 or more occurrences''.
\item {\tt +} after a non-terminal means ``1 or more occurrences''.
\item $t \theta$ denotes the result of applying substitution $\theta$ to
  $t$.  We represent substitutions as functions from variables to
  expressions.
\end{itemize}

\myparag{Well-formedness requirements on programs}

\begin{enumerate}

\item The top-level method in a program must be named {\tt main}.  It gets
  executed in an instance of the pre-defined {\tt process} class when the
  program starts.

\item Each label used in a {\tt receive} definition must be the label of
  some statement that appears in the same class as the {\tt receive}
  definition.

\item Invocations of methods defined using {\tt def} appear only in method
  call statements.  Invocations of methods defined using {\tt defun} appear
  only in method call expressions.
\end{enumerate}

\myparag{Constructs whose semantics is given by translation}

\begin{enumerate}

\item Constructors for all classes, and {\tt setup()} methods for process
  classes, are eliminated by translation into ordinary methods that assign
  to the fields of the objects.

\item A method call or field assignment that does not explicitly specify
  the target object is translated into a method call or field assignment,
  respectively, on {\tt self}.

\item An {\tt await} statement without an explicitly specified label---in
  other words, the associated label is the empty string---is translated
  into an {\tt await} statement with an explicitly specified label, by
  generating a fresh label name $\ell$, replacing the empty label in that
  {\tt await} statement with $\ell$, and inserting $\ell$ in every {\tt at}
  clause in the class containing the {\tt await} statement.

\item The Boolean operators {\tt and} and {\tt each} are eliminated as
  follows: {\tt $e_1$ and $e_2$} is replaced with {\tt not(not($e_1$) or
    not($e_2$))}, and {\tt each {\it iter} | $e$} is replaced with {\tt
    not(some {\it iter} | not($e$))}.

\item An aggregation is eliminated by translation into a comprehension
  followed by a {\tt for} loop that iterates over the set returned by the
  comprehension.  The {\tt for} loop updates an accumulator variable using
  the aggregation operator.

\item Iterators containing tuple patterns are rewritten as iterators
  without tuple patterns, as follows.

\begin{itemize}

\item Consider the existential quantification {\tt some \conf{\!}($e_1,
    \ldots, e_n$) in $s$ | $b$}.  Let $x$ be a fresh variable.  Let
  $\theta$ be the substitution that replaces $e_i$ with {\tt
    select($x$,$i$)} for each $i$ such that $e_i$ is a variable not
  prefixed with ``{\tt =}''.  Let $\{j_1,\ldots,j_m\}$ contain the indices
  of the constants and the variables prefixed with ``{\tt =}'' in {\tt
    ($e_1, \ldots ,e_n$)}.  Let $\bar e_j$ denote $e_j$ after removing the
  ``{\tt =}'' prefix, if any.  The quantification is rewritten as {\tt some
    $x$ in $s$ | isTuple($x$) and len($x$) is $n$ and (select($x$,$j_1$),
    $\ldots$, select($x$,$j_m$)) is ($\bar e_{j_1}$, $\ldots$, $\bar
    e_{j_m}$) and $b\theta$}.

\item Consider the loop {\tt for ($e_1,\ldots,e_n$) in $e$ : $s$}.  Let $x$ and
 $S$ be fresh variables.  Let $\{i_1,\ldots,i_k\}$ contain the indices
  in {\tt ($e_1, \ldots ,e_n$)} of variables not prefixed with ``{\tt =}''.
  Let $\theta$ be the substitution that replaces $e_i$ with {\tt
    select($x$,$i$)} for each $i$ in $\{i_1,\ldots,i_k\}$.  Let
  $\{j_1,\ldots,j_m\}$ contain the indices in {\tt ($e_1, \ldots ,e_n$)} of
  the constants and the variables prefixed with ``{\tt =}''.  Let $\bar
  e_j$ denote $e_j$ after removing the ``{\tt =}'' prefix, if any.  Note
  that $e$ may denote a set or sequence, and duplicate bindings for the
  tuple of variables $(e_{i_1},\ldots,e_{i_k})$ are filtered out if $e$ is
  a set but not if $e$ is a sequence.  The loop is rewritten as the code in
  Figure \ref{fig:elim-tuple-from-for-loop}.
 
\begin{figure*}[htb]
\setstretch{0.9}
\begin{alltt}
      \(S\) := \(e\)
      if isinstance(\(S\),set):
        \(S\) := \{ \(x\) : \(x\) in \(S\) | isTuple(\(x\)) and len(\(x\)) is \(n\)
              and (select(\(x\),\(j\sb{1}\)), \(\ldots\), select(\(x\),\(j\sb{m}\))) is (\(\bar{e}\sb{j\sb{1}}\), \(\ldots\), \(\bar{e}\sb{j\sb{m}}\)) \}
        for \(x\) in \(S\): 
          \(s\theta\)
      else: {\rm \commentMark \(S\) is a sequence}
        for \(x\) in \(S\): 
          if (isTuple(\(x\)) and len(\(x\)) is \(n\) 
              and (select(\(x\),\(j\sb{1}\)), \(\ldots\), select(\(x\),\(j\sb{m}\))) is (\(\bar{e}\sb{j\sb{1}}\), \(\ldots\), \(\bar{e}\sb{j\sb{m}}\)): 
            \(s\theta\)
          else: 
            skip
\end{alltt}
\caption{Translation of {\tt for} loop to eliminate tuple pattern.}
\label{fig:elim-tuple-from-for-loop}
\end{figure*}

\end{itemize}

\item Comprehensions in which some variables are prefixed with {\tt =} are
  translated into comprehensions without such prefixing.  Specifically, for
  a variable {\tt x} prefixed with {\tt =} in a comprehension, replace
  occurrences of {\tt =x} in the comprehension with occurrences of a fresh
  variable {\tt y}, and add the conjunct {\tt y is x} to the Boolean
  condition.

\item Comprehensions are statically eliminated as follows.  The
  comprehension {\tt $x$ := \{ $e$ | $x_1$ in $e_1$, $\ldots$, $x_n$
    in $e_n$ | $b$ \}}, where each $x_i$ is a pattern, is replaced with
\begin{alltt}\setstretch{0.9}
\(x\) := new set
for \(x\sb{1}\) in \(e\sb{1}\):
  ...
    for \(x\sb{n}\) in \(e\sb{n}\):
      if \(b\):
        \(x\).add(\(e\))
\end{alltt}

\item Wildcards are eliminated from tuple patterns by replacing each
  occurrence of wildcard with a fresh variable.

\item Remote method invocation, i.e., invocation of a method on another
  process after that process has been started, is translated into message
  communication.

\end{enumerate}

\myparag{Notes}

\begin{enumerate}

\item \ClassName\ must include {\tt process}.  {\tt process} is a
  pre-defined class; it should not be defined explicitly.  {\tt process}
  has fields {\tt sent} and {\tt received}, and it has a method {\tt
    start}.  %
      
\item The grammar allows {\tt receive} definitions to appear in classes
  that do not extend {\tt process}, but such {\tt receive} definitions are
  useless, so it would be reasonable to make them illegal.

\item The grammar does not allow labels on statements other than {\tt
    await}.  A label $\ell$ on a statement $s$ other than {\tt await} is
  treated as syntactic sugar for label $\ell$ on {\tt await true : skip}
  followed by statement $s$.

\item \ClassName\ must include {\tt set} and {\tt sequence}.  Sets and
  sequences are treated as objects, because they are mutable.  These are
  predefined classes that should not be defined explicitly.  Methods of
  {\tt set} include {\tt add}, {\tt del}, {\tt contains}, {\tt min}, {\tt
    max}, and {\tt size}.  Methods of {\tt sequence} include {\tt add}
  (which adds an element at the end of the sequence), {\tt contains}, and
  {\tt length}.  We give the semantics explicitly for a few of these
  methods; the others are handled similarly.

\item Tuples are treated as immutable values, not as mutable objects.

\item All expressions are side-effect free.  For simplicity, we treat
  quantifications as expressions, so existential quantifications do not
  have the side-effect of binding variables to a witness.  Such existential
  quantifications could be added as a new form of statement.

\item Object creation and comprehension are statements, not expressions,
  because they have side-effects.  Comprehension has the side-effect of
  creating a new {\tt set}.

\item \Parameter\ must include {\tt self}.  The values of method
  parameters cannot be updated (e.g., using assignment statements).  For
  brevity, local variables of methods are omitted from the core language.
  Consequently, assignment is allowed only for instance variables.

\item Semantically, the {\tt for} loop copies the contents of a (mutable)
  sequence or set into an (immutable) tuple before iterating over it, to
  ensure that changes to the sequence or set by the loop body do not affect
  the iteration.  An implementation could use optimizations to achieve this
  semantics without copying when possible.

\item For brevity, among the standard arithmetic operations ({\tt +}, {\tt
    -}, {\tt *}, etc.), we include only one representative operation in the
  abstract syntax and semantics; others are handled similarly.

\item The semantics below does not model real-time, so timeouts in {\tt
    await} statements are simply allowed to occur non-deterministically.

\item We omit the concept of node (process location) from the semantics,
  and we omit the node argument of the constructor when creating instances
  of process classes, because process location does not affect other
  aspects of the semantics.

\item We omit {\tt configure handling} statements from the syntax.  The
  semantics is for {\tt configure handling = all}.  Semantics for other
  {\tt configure handling} options can easily be added.

\item To support initialization of a process by its parent, a process can
  access fields of another process and invoke methods on another process
  before the latter process is started.

\item We require that all messages be tuples.  This is an inessential
  restriction; it slightly simplifies the specification of pattern matching
  for matching messages against patterns.

\item A process's {\tt sent} sequence contains pairs of the form $(m,d)$,
  where $m$ is a message sent by the process to destination $d$.  A
  process's {\tt received} sequence contains pairs of the form $(m,s)$,
  where $m$ is a message received by the process from sender $s$.  
\end{enumerate}

\mysection{Semantic Domains}
\label{sec:domains}

The semantic domains are defined in Figure \ref{fig:domains}.

\myparag{Notation}
\begin{itemize}
\item $D^*$ contains finite sequences of values from domain $D$.  
\item $\Set{D}$ contains finite sets of values from domain $D$.
\item $D1 \pfn D2$ contains partial functions from $D_1$ to $D_2$.
  $\dom(f)$ is the domain of a partial function $f$.
\end{itemize}

\newcommand{\negspc}{\conf{\!\!\!}}
\begin{figure}[htb]
\setstretch{0.9}
\conf{\hspace*{-5em}}
\begin{eqnarray*}
  \Bool \negspc&=&\negspc \set{{\tt true}, {\tt false}}\\
  \Int \negspc&=&\negspc ... \\
  \ProcessAddress \negspc&=&\negspc ...\\
  \NonProcessAddress \negspc&=&\negspc ...\\
  \Address \negspc&=&\negspc \ProcessAddress \union \NonProcessAddress\\
  \Tuple \negspc&=&\negspc \Val^*\\
  \Val \negspc&=&\negspc \Bool \union \Int \union \Address \union \Tuple\\
  \SetOfVal \negspc&=&\negspc \Set{\Val}\\
  \SeqOfVal \negspc&=&\negspc \Val^*\\
  \Object \negspc&=&\negspc (\Field \pfn \Val) \union \SetOfVal \union \SeqOfVal\\
  \HeapType \negspc&=&\negspc \Address \pfn \ClassName\\
  \LocalHeap \negspc&=&\negspc \Address \pfn \Object\\
  \Heap \negspc&=&\negspc \ProcessAddress \pfn \LocalHeap\\
  \ChannelStates \negspc&=&\negspc \ProcessAddress \times \ProcessAddress \conf{\\ & &} \pfn \Tuple^*\\
  \MsgQueue \negspc&=&\negspc (\Tuple \times \ProcessAddress)^*\\
  \State \negspc&=&\negspc (\ProcessAddress \pfn \Statement)\conf{\\
  \negspc& &\negspc} \times \HeapType \times \Heap \times \ChannelStates\\
  \negspc& &\negspc \times (\ProcessAddress \pfn \MsgQueue)\\
\end{eqnarray*}
\caption{Semantic domains.  Ellipses are used for semantic domains of
  primitive values whose details are standard or unimportant.}
  \label{fig:domains}
\end{figure}

\myparag{Notes}
\begin{itemize}
\item We require that $\ProcessAddress$ and $\NonProcessAddress$ be disjoint.
\item For $a\in \ProcessAddress$ and $h\in\Heap$, $h(a)$ is the local heap
  of process $a$.  For $a\in \Address$ and $ht\in\HeapType$, $ht(a)$ is the
  type of the object with address $a$.  For convenience, we use a single
  (global) function for $\HeapType$ in the semantics, even though the
  information in that function is distributed in the same way as the heap
  itself in an implementation.
\item The $\MsgQueue$ associated with a process by the last component of a
  state contains messages, paired with the sender, that have arrived at the
  process but have not yet been handled by matching {\tt receive}
  definitions.
\end{itemize}

\mysection{Extended Abstract Syntax}
\label{sec:extended-syntax}

Section \ref{sec:syntax} defines the abstract syntax of programs that can
be written by the user.  Figure \ref{fig:extended-syntax} extends the
abstract syntax to include additional forms into which programs may evolve
during evaluation.  Only the new productions are shown here; all of the
productions given above carry over unchanged.

\begin{figure}[htb]
\setstretch{0.9}
\begin{ctabbing}
\Expression\ ::= \= {\it Address}\\
\> {\it Address}.\Field\\
\\
\Statement\ ::= {\tt for} {\it Variable} {\tt intuple} \Tuple: \Statement
\end{ctabbing}
  \caption{Extensions to the abstract syntax.}
  \label{fig:extended-syntax}
\end{figure}

The statement {\tt for $v$ intuple $t$:~$s$} iterates over the elements of
tuple $t$, in the obvious way.

\mysection{Evaluation Contexts}

Evaluation contexts, also called reduction contexts, are used to identify
the next part of an expression or statement to be evaluated.  An evaluation
context is an expression or statement with a hole, denoted {\tt []}, in
place of the next sub-expression or sub-statement to be evaluated.
Evaluation contexts are defined in Figure \ref{fig:eval-context}.

\begin{figure}[htb]
\setstretch{0.9}
\begin{ctabbing}
\Val\ ::= \=
 \Literal\\
\> \Address\\
\> (\Val*)
\\
\EC\ ::= \=
 []\\
\> (\Val*,\EC,\Expression*)\\
\> \EC.\MethodName(\Expression*)\\
\> \Address.\MethodName(\Val*,\EC,\Expression*)\\
\> \UnaryOp(\EC)\\
\> \BinaryOp(\EC,\Expression)\\
\> \BinaryOp(\Val,\EC)\\
\> {\tt isinstance}(\EC,\ClassName)\\
\> {\tt or}(\EC,\Expression)\\
\> {\tt some} \Pattern\ {\tt in} \EC\ {\tt |} \Expression\\
\> \EC.\Field\ := \Expression\\
\> \Address.\Field\ := \EC\\
\> \InstanceVariable\ := \EC\\
\> \EC\ ; \Statement\\
\> {\tt if} \EC{\tt :} \Statement\ {\tt else:} \Statement\\
\> {\tt for} \InstanceVariable\ {\tt in} {\it \EC}{\tt :} \Statement\\
\> {\tt for} \InstanceVariable\ {\tt intuple} \Tuple{\tt :} \EC\\
\> {\tt send} \EC\ {\tt to} \Expression\\
\> {\tt send} \Val\ {\tt to} \EC\\
\> {\tt await} \Expression\ {\tt :} \Statement\ \AnotherAwaitClause*\conf{\\ \hspace*{3.5em}} {\tt timeout} \EC
\end{ctabbing}
   \caption{Evaluation contexts.}
   \label{fig:eval-context}
 \end{figure}

\mysection{Transition Relations}
\label{sec:transition}

The transition relation for expressions has the form $ht:h \vdash e \ra e'$,
where $e$ and $e'$ are expressions, $ht\in\HeapType$, and $h\in\LocalHeap$.
The transition relation for statements has the form $\sigma \ra \sigma'$
where $\sigma\in\State$ and $\sigma'\in\State$.  

Both transition relations are implicitly parameterized by the program,
which is needed to look up method definitions and configuration
information.  The transition relation for expressions is defined in Figure
\ref{fig:transition-expr}.  The transition relation for statements is
defined in Figures
\ref{fig:transition-statement-1}--\ref{fig:transition-statement-2}.

\myparag{Notation and auxiliary functions}
\begin{itemize}
\item In the transition rules, $a$ matches an address; 
  $v$ matches a value (i.e., an element of $\Val$);
  and $\ell$ matches a label.

\item For an expression or statement $e$, $e[x := y]$ denotes $e$ with all
  occurrences of $x$ replaced with $y$.

\item A function matches the pattern $f[x \ra y]$ iff $f(x)$ equals $y$.  For example, in transition rules for statements, a function $P$ in $\ProcessAddress \pfn \Statement$ matches $P[a \ra s]$ if $P$ maps process address $a$ to statement $s$.

\item For a function $f$, $f[x := y]$ denotes the function that is the same
  as $f$ except that it maps $x$ to $y$.

\item $\emptyfn$ denotes the empty partial function, i.e., the partial
  function whose domain is the empty set.

\item For a (partial) function $f$, $f \ominus a$ denotes the function that
  is the same as $f$ except that it has no mapping for $a$.

\item Sequences are denoted with angle brackets, e.g., $\seq{0,1,2} \in
  \Int^*$.

\item $s@t$ is the concatenation of sequences $s$ and $t$.

\item $\first(s)$ is the first element of sequence $s$.

\item $\rest(s)$ is the sequence obtained by removing the first element of $s$.

\item $\length(s)$ is the length of sequence $s$.

\item $\extends(c_1,c_2)$ holds iff class $c_1$ is a descendant of class
  $c_2$ in the inheritance hierarchy.

\item For $c\in\ClassName$, 
  $\new(c)$ returns a new instance of $c$.  
  \begin{displaymath}
    \begin{array}{@{}l@{}}
    \new(c) = \conf{\\
    ~~~~}\left\{\begin{array}{@{}ll@{}} 
        \emptySet & \mbox{if } c={\tt set}\\
        \emptyseq & \mbox{if } c={\tt sequence}\\
        \emptyfn & \mbox{otherwise}
      \end{array}\right.
    \end{array}
  \end{displaymath}

\item For $m\in \MethodName$ and $c \in \ClassName$, the relation
  $\methodDef(c,\linebreak[0] m, {\it def})$ holds iff (1) class $c$ defines
  method $m$, and {\it def} is the definition of $m$ in $c$, or (2) $c$
  does not define $m$, and {\it def} is the definition of $m$ in the
  nearest ancestor of $c$ in the inheritance hierarchy that defines $m$.

\item For $h, \bar h, \bar h'\in\LocalHeap$ and $ht, ht'\in \HeapType$ and
  $v, \bar v\in\Val$, the relation $\iscopy(v, h, \bar h, ht,$ $\bar v, \bar h', ht')$ holds iff
  (1) $v$ is a value in a process with local heap $h$, i.e., addresses in
  $v$ are evaluated with respect to $h$, (2) $\bar v$ is a copy of $v$ for a
  process whose local heap was $\bar h$ before $v$ was copied into it and
  whose local heap is $\bar h'$ after $v$ is copied into it, i.e., $\bar v$ is
  the same as $v$ except that, instead of referencing objects in $h$, it
  references newly created copies of those objects in $\bar h'$, and (3)
  $\bar h'$ and $ht'$ are versions of $\bar h$ and $ht$ updated to reflect
  the creation of those objects.  As an exception, because process
  addresses are used as global identifiers, process addresses in $v$ are
  copied unchanged into $\bar v$, and new copies of process objects are not
  created.  We give auxiliary definitions and then a formal definition
  of $\iscopy$.  

  For $v\in\Val$, let $\addrs(v,h)$ denote the set of addresses that appear
  in $v$ or in any objects or values reachable from $v$
  with respect to local heap $h$; formally,
  \begin{displaymath}
    \tran{\hspace*{3em}}
    \begin{array}[t]{@{}l@{}}
      a \in \addrs(v,h) \myiff \\
      \spce (v \in \Address \land v=a)\\
      \spce {}\lor (v \in \dom(h) \land h(v) \in \Field\pfn\Val
      \conf{\\ ~~~~~~~~~}\land (\exists f \in \dom(h(v)).\; a \in \addrs(h(v)(f), h)))\\
      \spce {}\lor (v \in dom(h) \land h(v) \in \SetOfVal\union\SeqOfVal \conf{\\ ~~~~~~~~~}\land (\exists v' \in
      h(v).\; a \in \addrs(v', h)))\\
      \spce {}\lor (\exists v_1,\ldots,v_n\in\Val.\; v = (v_1,\ldots, v_n)
      \conf{\\ ~~~~~~~~~} \land \exists i\in [1..n].\; a \in \addrs(v_i, h))
    \end{array}
  \end{displaymath}
  For $v,\bar v\in\Val$ and $f \in \Address \pfn \Address$, the relation $\subst(v,\bar v,f)$
  holds iff $v$ is obtained from $\bar v$ by replacing each occurrence of an
  address $a$ in $\dom(f)$ with $f(a)$ (informally, $f$ maps addresses of
  new objects in $\bar v$ to addresses of corresponding old objects in $v$);
  formally,
  \begin{displaymath}
    \tran{\hspace*{3em}\subst(v,\bar v,f) \myiff }
    \begin{array}[t]{@{}l@{}}
    \conf{\subst(v,\bar v,f) = \\}
    \conf{~~~~} (v \in \Bool\union\Int\union(\Address\setminus\dom(f)) \land \bar v=v)\\
    \conf{~~~~} \lor (v \in \dom(f) \land f(\bar v)=v)\\
\conf{~~~~} \lor (\exists v_1, \ldots, v_n, \bar v_1, \ldots, \bar v_n.\; \conf{\\ ~~~~~~~~~~~}
      v=(v_1,\ldots,v_n) \land \bar v=(\bar v_1,\ldots,\bar v_n)\\ 
      \tran{\spce}\conf{~~~~~~~~~~} 
      \land (\forall i \in [1..n].\; subst(v_i, \bar v_i, f)))
    \end{array}
  \end{displaymath}
  Similarly, for $o,\bar o \in \Object$ and $f \in \Address \pfn \Address$,
  the relation $\subst(o,\bar o,f)$ holds iff $o$ is obtained from $\bar o$ by replacing
  each occurrence of an address $a$ in $\dom(f)$ with $f(a)$.  For sets $S$
  and $S'$, let $S \bijection S'$ be the set of bijections between $S$ and
  $S'$.

  Finally, $\iscopy$ is defined as follows (intuitively, $A$ contains the
  addresses of the newly allocated objects):
  \begin{displaymath}
    \tran{\hspace*{3em}}
    \begin{array}[t]{@{}l@{}}
      \iscopy(v, h, \bar h, ht, \bar v, \bar h ', ht') \myiff \\
      \spce \exists A \subset \NonProcessAddress.\; 
      \conf{\\ \spce} \exists f \in A \bijection (\addrs(v,h)\setminus\ProcessAddress).\\
       \spce\sci A \intersect\dom(ht)=\emptyset
       \conf{\\ \spce\sci} {}\land \dom(ht')=\dom(ht) \union A
       \conf{\\ \spce\sci} {}\land \dom(\bar h') = \dom(\bar h) \union A\\
       \spce\sci {}\land (\forall a \in \dom(ht).\; ht'(a) = ht(a))
\conf{\\ \spce\sci} {}\land (\forall a \in \dom(\bar h).\; \bar h'(a) = \bar h(a))\\
       \spce\sci {}\land (\forall a \in A.\; ht'(a) = ht(f(a)) \conf{\\ \spce\sci\sci\sci} {}\land \subst(h(a), \bar h'(a), f))
    \end{array}
  \end{displaymath}

\newcommand{\matchesDefLbl}{{\it matchesDefLbl}}
\newcommand{\findSubstPat}{{\it findSubstPat}}
\newcommand{\findSubst}{{\it findSubst}}
\newcommand{\bound}{{\it bound}}
\newcommand{\vars}{{\it vars}}

\item For $m\in\Val$, $a\in\ProcessAddress$, $\ell \in\Label$, $h\in\LocalHeap$, and a {\tt receive} definition $d$, if message $m$ can be received from $a$ at label $\ell$ by a process with local heap $h$ using {\tt receive} definition $d$, then $\matchRcvDef(m, a, \ell, h, d)$ returns the appropriately instantiated body of $d$.  

We first define some auxiliary relations and functions.  The relation $\matchesDefLbl(d,\ell)$ holds iff {\tt receive} definition $d$ either lacks an {\tt at} clause or has an {\tt at} clause that includes $\ell$.  $\bound(P)$ returns the set of variables that appear in pattern $P$ prefixed with ``{\tt =}''.  $\vars(P)$ returns the set of variables that appear in $P$.  $\findSubstPat(m, a, h, P~{\tt from}~x)$ returns the substitution $\theta$ with domain $\vars(P)\union\set{x}$ such that $m=P\theta \land \theta(x)=a \land (\forall y \in \bound(P).\; \theta(y)=h(y))$, if any, otherwise it returns $\bot$.  $\findSubst(m, a, h, d)$ returns $\findSubstPat(m, a, h, P~{\tt from}~x)$ for the first {\tt receive} pattern $P~{\tt from}~x$ in $d$ such that $\findSubstPat(m, a, h,$ $P~{\tt from}~x)\ne \bot$, if any, otherwise it returns $\bot$.  

If $\matchesDefLbl(d, \ell) \land \findSubst(m, a, h, d)\ne \bot$, then $\matchRcvDef(m, a, \ell, h, d)$ returns $s\theta$, where $s$ is the body of $d$ (i.e., the statement that appears in $d$) and $\theta = \findSubst(m, a, h, d)$, otherwise it returns $\bot$.

\item For $m\in\Val$, $a\in\ProcessAddress$, $\ell \in\Label$,
  $c\in\ClassName$, and $h\in\LocalHeap$, if message $m$ can be received
  from $a$ at label $\ell$ in class $c$ by a process with local heap $h$,
  then $\rcvAtLabel((m, a), \ell, c, h)$ returns a set of statements that should be
  executed when receiving $m$ in that context.

  Specifically, if class $c$ contains a {\tt receive} definition $d$ such
  that $\matchRcvDef(m, a, \ell, h, d)$ is not $\bottom$, then, letting
  $d_1,\ldots,d_n$ be the {\tt receive} definitions in $c$ such that
  $\matchRcvDef(m, a,$ $\ell, h, d_i)$ is not $\bottom$, and letting
  $s_i=\matchRcvDef(m, a, \ell, h, d_i)$, $\rcvAtLabel((m, a), \ell, c, h)$
  returns $\set{s_1, \ldots, s_n}$.  Otherwise, $\rcvAtLabel((m, a), \ell,$
  $c, h)$ returns $\emptyset$. %

\end{itemize}

\begin{figure*}[tbp]
\setstretch{0.9}
\begin{displaymath}
\begin{array}{@{}l@{}}
\commentS{field access}\\
ht:h \vdash a.f \ra h(a)(f) \spce \IF a \in \dom(h) \land f \in \dom(h(a))\\
\\
\commentS{invoke method in user-defined class}\\
ht:h \vdash a{\tt .}m(v_1,\ldots,v_n) \ra e[{\tt self}:=a, x_1:=v_1, \ldots, x_n:=v_n]\\
\sci \IF a \in \dom(h) \land \methodDef(ht(a), m, {\tt defun}~m(x_1,\ldots,x_n)~e)\\
\\
\commentS{invoke method in pre-defined class (representative examples)}\\
ht:h \vdash a{\tt .contains(}v_1{\tt )} \ra {\tt true}
\spce \IF a \in \dom(h) \land ht(a)={\tt set} \land v_1 \in h(a)\\
ht:h \vdash a{\tt .contains(}v_1{\tt )} \ra {\tt false}
\spce \IF a \in \dom(h) \land ht(a)={\tt set} \land v_1 \not\in h(a)\\
\\
\commentS{unary operations}\\
ht:h \vdash {\tt not(true)} \ra {\tt false}\\
ht:h \vdash {\tt not(false)} \ra {\tt true}\\
ht:h \vdash {\tt isTuple(}v{\tt )} \ra {\tt true} \spce 
\mbox{if $v$ is a tuple}\\
ht:h \vdash {\tt isTuple(}v{\tt )} \ra {\tt false} \spce 
\mbox{if $v$ is not a tuple}\\
ht:h \vdash {\tt len(}v{\tt )} \ra n  \spce 
\mbox{if $v$ is a tuple with $n$ components}\\
\\
\commentS{binary operations}\\
ht:h \vdash {\tt is(}v_1,v_2{\tt )} \ra {\tt true} \spce 
\mbox{if $v_1$ and $v_2$ are the same (identical) value}\\
\\
ht:h \vdash {\tt plus(}v_1,v_2{\tt )} \ra v_3   \spce 
\IF v_1\in\Int \land v_2\in\Int \land v_3 = v_1 + v_2\\
\\
ht:h \vdash {\tt select(}v_1,v_2{\tt )} \ra v_3\\
\sci \IF v_2\in\Int \land v_2>0 \land
\mbox{($v_1$ is a tuple with at least $v_2$ components)} \tran{\\ \sci\sci{}} \land
\mbox{($v_3$ is the $v_2$'th component of $v_1$)}\\
\\
\commentS{isinstance}\\
ht:h \vdash {\tt isinstance(}a, c{\tt )} \ra {\tt true} \spce \IF ht(a)=c\\
ht:h \vdash {\tt isinstance(}a, c{\tt )} \ra {\tt false} \spce \IF ht(a)\ne c\\
\\
\commentS{disjunction}\\
ht:h \vdash {\tt or(true}, e{\tt )} \ra {\tt true}\\
ht:h \vdash {\tt or(false}, e{\tt )} \ra e\\
\\
\commentS{existential quantification}\\
ht:h \vdash {\tt some}~x~{\tt in}~a~|~e ~~\ra~~ e[x:=v_1]~{\tt or }~\cdots~{\tt or}~e[x:=v_n]\\
\sci \IF (ht(a)={\tt sequence} \land h(a)=\seq{v_1,\ldots,v_n}) \tran{\\ \sci\sci{}}
    \lor (ht(a)={\tt set} \land \seq{v_1,\ldots,v_n}~\mbox{is a linearization of}~h(a))
\end{array}
\end{displaymath}
  \caption{Transition relation for expressions.}
  \label{fig:transition-expr}
\end{figure*}

\begin{figure*}[tbp]
\setstretch{0.9}
{\tran{\small}
\begin{displaymath}
\begin{array}{@{}l@{}}
\commentS{field assignment}\\
\tuple{P[a \ra a'{\tt .}f \;{\tt :=}\; v], ht, h[a \ra ha[a' \ra o]], ch, mq}\\
{}\ra 
\tuple{P[a := {\tt skip}], ht, h[a := ha[a' := o[f := v]]], ch, mq}\\
\\
\commentS{object creation}\\
\tuple{P[a \ra a'{\tt .}f \;{\tt :=}\; {\tt new}~c], ht, h[a \ra ha[a' \ra o]], ch, mq}\\
{}\ra \tuple{P[a := {\tt skip}], ht[a' := c], h[a := ha[a' := o[f := a_c], a_c := \new(c)]], ch, mq}\\
\sci \IF a_c\not\in\dom(ht) 
\land a_c \in \Address \land (a_c \in\ProcessAddress \myiff \extends(c,{\tt process}))\\
\\
\commentS{sequential composition}\\
\tuple{P[a \ra {\tt skip;}\, s], ht, h, ch, mq} \ra
\tuple{P[a := s], ht, h, ch, mq}\\
\\
\commentS{conditional statement}\\
\tuple{P[a \ra {\tt if}~{\tt true:}~s_1~{\tt else:}~s_2], ht, h, ch, mq} \ra
\tuple{P[a := s_1], ht, h, ch, mq}\\
\\
\tuple{P[a \ra {\tt if}~{\tt false:}~s_1~{\tt else:}~s_2], ht, h, ch, mq} \ra
\tuple{P[a := s_2], ht, h, ch, mq}\\
\\
\commentS{for loop}\\
\tuple{P[a \ra {\tt for}~x~{\tt in}~a'{\tt :}~s], ht, h, ch, mq} \ra
\tuple{P[a := {\tt for}~x~{\tt intuple}~{\tt (}v_1,\ldots,v_n{\tt ):}~s], ht, h, ch, mq}\\
\sci \IF ((ht(a)={\tt sequence} \land h(a)(a')=\seq{v_1,\ldots,v_n})\tran{\\ \sci\spce{}} \lor
    (ht(a)={\tt set} \land \seq{v_1,\ldots,v_n}~\mbox{is a linearization of}~h(a)(a')))\\
\\
\tuple{P[a \ra {\tt for}~x~{\tt intuple}~{\tt (}v_1,\ldots,v_n{\tt ):}~s], ht, h, ch, mq}\\
{}\ra \tuple{P[a := s[x:=v_1]; {\tt for}~x~{\tt intuple}~{\tt (}v_2,\ldots,v_n{\tt ):}~s], ht, h, ch, mq}\\
\\
\tuple{P[a \ra {\tt for}~x~{\tt intuple}~{\tt ():}~s], ht, h, ch, mq} \ra
\tuple{P[a := {\tt skip}], ht, h, ch, mq}\\
\\
\commentS{while loop}\\
\tuple{P[a \ra {\tt while}~e{\tt :}~s], ht, h, ch, mq} \ra
\tuple{P[a := {\tt if}~e{\tt :}~{\tt (}s{\tt ;}~{\tt while}~e{\tt :}~s{\tt )}~{\tt else:}~{\tt skip}], ht, h, ch, mq}\\
\\
\commentS{invoke method in user-defined class}\\
\tuple{P[a \ra a'{\tt .}m(v_1,\ldots,v_n)], ht, h, ch, mq}\\
{} \ra
\tuple{P[a := s[{\tt self}:=a, x_1:=v_1, \ldots, x_n:=v_n]], ht, h, ch, mq}\\
\sci\IF a'\in\dom(h(a))\\
\sci~~ {}\land ht(a')\not\in\set{{\tt process}, {\tt set}, {\tt sequence}}
\land \methodDef(ht(a'), m, {\tt def}~m{\tt (}x_1,\ldots,x_n{\tt )}~s)\\
\\
\commentS{invoke method in pre-defined class (representative examples)}\\
\\
\commentS{{\tt process.start} allocates a local heap and {\tt sent} and {\tt received} sequences
for the new process,}\\
\commentS{and moves the started process to the new local heap.}\\
\tuple{P[a \ra a'{\tt .start()}], ht, h[a \ra ha[a' \ra o], ch, mq}\\
{}\ra 
\begin{array}[t]{@{}l@{}}
\ltuple{P[a := {\tt skip}, a' := a'{\tt .run()}], ht[a_s := {\tt sequence},a_r := {\tt sequence}],}\\
~~\rtuple{h[a := ha \ominus a', a' := \emptyfn[a' \ra o[{\tt sent} := a_s, {\tt received} := a_r], a_r := \emptyseq, a_s := \emptyseq]], ch, mq}
\end{array}
\\
\sci \IF \extends(ht(a'),{\tt process}) 
\land \mbox{($ht(a')$ inherits {\tt start} from {\tt process})}
\land a_r\not\in\dom(ht) \land a_s\not\in\dom(ht)\\
\sci~~ {} \land a_r \in \NonProcessAddress
\land a_s \in \NonProcessAddress\\
\\
\tuple{P[a \ra a'{\tt .add(}v_1{\tt )}], ht, h[a \ra ha], ch, mq}\\
{} \ra
\tuple{P[a := {\tt skip}], ht, h[a := ha[a' := ha(a')\union\set{v_1}]], ch, mq}\\
\sci \IF a'\in\dom(ha) \land ht(a')={\tt set} \\
\\
\tuple{P[a \ra a'{\tt .add(}v_1{\tt )}], ht, h[a \ra ha], ch, mq}\\
{} \ra
\tuple{P[a := {\tt skip}], ht, h[a := ha[a' := ha(a')@\seq{v_1}]], ch, mq}\\
\sci \IF a'\in\dom(ha) \land ht(a')={\tt sequence}
\end{array}
\end{displaymath}}
  \caption{Transition relation for statements, Part 1.}
  \label{fig:transition-statement-1}
\end{figure*}

\begin{figure*}[tbp]
 \setstretch{0.9}
{\tran{\small}
\begin{displaymath}
\begin{array}{@{}l@{}}
\commentS{send a message to one process.  create copies of the message for the sender's
{\tt sent} sequence}\\
\commentS{and the receiver.}\\
\tuple{P[a \ra {\tt send}~v~{\tt to}~a_2], ht, h[a \ra ha, a_2\ra ha_2], ch, mq}\\
{}\ra
(P[a := {\tt skip}], ht'', 
h[a := ha'[a_s := ha(a_s)@\seq{(v_1,a_2)}], a_2 := ha'_2],\\
\sci\spce ch[\tuple{a,a_2} := ch(\tuple{a,a_2})@\seq{v_2}], mq)\\
\sci \IF a_2 \in \ProcessAddress \land a_s = ha(a)({\tt sent})
\land \iscopy(v,ha,ha,ht,v_1,ha',ht') \\
\sci\sci{}\land \iscopy(v,ha',ha_2,ht',v_2,ha'_2,ht'')\\
\\
\commentS{send to a set of processes}\\
\tuple{P[a \ra {\tt send}~v~{\tt to}~a'], ht, h[a \ra ha], ch, mq}\\
{}\ra \tuple{P[a := {\tt for}~x~{\tt in}~a'{\tt :}~{\tt send}~v~{\tt to}~x], ht, h[a := ha[a_s := ha(a_s)@\seq{(v,a')}]], ch, mq}\\
\sci \IF ht(a')={\tt set}
\land a_s = ha(a)({\tt sent})
\land \mbox{($x$ is a fresh variable)}\\
\\
\commentS{message reordering}\\
\tuple{P, ht, h, ch[\tuple{a,a'} \ra q], mq} \ra
\tuple{P, ht, h, ch[\tuple{a,a'} := q'], mq}\\
\sci \IF~\mbox{(channel order is {\tt unordered} in the program configuration)}
\land \mbox{($q'$ is a permutation of $q$)}\\
\\
\commentS{message loss}\\
\tuple{P, ht, h, ch[\tuple{a,a'} \ra q], mq} \ra
\tuple{P, ht, h, ch[\tuple{a,a'} := q'], mq}\\
\sci \IF~\mbox{(channel reliability is {\tt unreliable} in the program configuration)}
\land \mbox{($q'$ is a subsequence of $q$)}\\
\\
\commentS{arrival of a message from process $a$ at process $a'$.  remove
  message from channel, and append}\\
\commentS{(message, sender) pair to message queue.}\\
\tuple{P, ht, h, ch[\tuple{a,a'} \ra q], mq}\\
{}\ra \tuple{P, ht, h, ch[\tuple{a,a'} := \rest(q)], mq[a' := mq(a')@\seq{\tuple{\first(q), a}}]}\\
\sci \IF %
\length(q) > 0\\
\\
\commentS{handle a message at a yield point.  remove the (message, sender) pair from the message}\\
\commentS{queue, append a copy to the {\tt received} sequence, and prepare
  to run matching receive}\\
\commentS{handlers associated with $\ell$, if any.  $s$ has a label hence must be {\tt await}.}\\
\tuple{P[a \ra \ell\ s], ht, h[a \ra ha], ch, mq[a \ra q]}\\
\ra \tuple{P[a := s'[{\tt self} := a]; \ell\ s], ht', h[a \ra ha'[a_r \ra ha(a_r)@\seq{copy}]], ch, mq[a := \rest(q)]}\\
\sci \IF \length(q)>0 \land a_r = ha(a)({\tt received}) \land 
\iscopy(\first(q), ha, ha, ht, copy, ha', ht') \\
\sci~~{}\land \rcvAtLabel(\first(q), \ell, ht(a), ha')=S \land
s'~\mbox{is a linearization of}~S
 \\
\\
\commentS{await without timeout clause}\\
\tuple{P[a \ra \ell\ {\tt await}~e_1{\tt :}s_1~{\tt or}~\cdots~{\tt or}~e_n{\tt :}s_n], ht, h, ch, mq} \ra 
\tuple{P[a := s_i], ht, h, ch, mq}\\
\sci \IF \length(mq(a))=0 \land i\in [1..n] \land h(a):ht \vdash e_i \ra {\tt true}\\
\\
\commentS{await with timeout clause, terminated by true condition}\\
\tuple{P[a \ra \ell\ {\tt await}~e_1{\tt :}s_1~{\tt or}~\cdots~{\tt or}~e_n{\tt :}s_n~{\tt timeout}~v{\tt :}s], ht, h, ch, mq} \ra 
\tuple{P[a := s_i], ht, h, ch, mq}\\
\sci \IF \length(mq(a))=0 \land i\in [1..n] \land h(a):ht \vdash e_i \ra {\tt true}\\
\\
\commentS{await with timeout clause, terminated by timeout (occurs non-deterministically)}\\
\tuple{P[a \ra \ell\ {\tt await}~e_1{\tt :}s_1~{\tt or}~\cdots~{\tt or}~e_n{\tt :}s_n~{\tt timeout}~v{\tt :}s], ht, h, ch, mq} \ra
\tuple{P[a := s], ht, h, ch, mq}\\
\sci \IF \length(mq(a))=0 \land h(a):ht \vdash e_1~ \ra {\tt false} \land \cdots \land h(a):ht \vdash e_n~ \ra {\tt false}\\
\\
\commentS{context rule for expressions}\\
\dfrac{h(a):ht \vdash e \ra e'}{\tuple{P[a \ra C[e]], ht, h, ch, mq} \ra \tuple{P[a := C[e']], ht, h, ch, mq}}\\
\\
\commentS{context rule for statements}\\
\dfrac{\tuple{P[a \ra s], ht, h, ch, mq} \ra \tuple{P[a := s'], ht', h', ch', mq'}}{\tuple{P[a \ra C[s]], ht, h, ch, mq} \ra \tuple{P[a := C[s']], ht', h', ch', mq'}}
\end{array}
\end{displaymath}}
  \caption{Transition relation for statements, Part 2.}
  \label{fig:transition-statement-2}
\end{figure*}

\mysection{Executions}
\label{sec:execution}

An execution is a sequence of transitions $\sigma_0 \ra \sigma_1 \ra
\sigma_2 \ra \cdots$ such that $\sigma_0$ is an initial state.  The
set of initial states is defined in Figure \ref{fig:init-state}.
\begin{figure}[htb]
 \setstretch{0.9}
\begin{displaymath}
  \begin{array}[t]{@{}l@{}}
  \Init = \\
  \{ \tuple{P, ht, h, ch, mq} \in \State \;|\; \\
  ~~
    \exists\; a_p \in \ProcessAddress, \\
\spce\begin{array}{@{}l@{}}
    a_r \in \NonProcessAddress, \\
    a_s \in \NonProcessAddress.\;\\
    a_r \ne a_s \\
    {}\land P = \emptyfn[a_p := a_p{\tt .main()}] \\
    {}\land\conf{\!} ht \conf{\!}=\conf{\!} \emptyfn[a_p \conf{\!}:=\conf{\!} {\tt process}, a_r \conf{\!}:=\conf{\!} {\tt sequence}, a_s \conf{\!}:=\conf{\!} {\tt sequence}]\\
    {}\land h = \emptyfn[a_p := ha]\\
    {}\land ch \conf{\!}=\conf{\!} (\lambda \tuple{a_1, a_2} \conf{\!}\in\conf{\!} \ProcessAddress \conf{\!}\times\conf{\!} \ProcessAddress.\tran{\;}\conf{\,} \emptyseq)\\
    {}\land mq = (\lambda a \in \ProcessAddress.\; \emptyseq)\\
    \mbox{where } 
    \begin{array}[t]{@{}l@{}}
    ha = \emptyfn[a_p:=o_p, a_r:=\emptyseq, a_s:=\emptyseq]\\
    o_p = \emptyfn[{\tt received} := a_r, {\tt sent} := a_s]
    \}
  \end{array}
    \end{array}
  \end{array}
\end{displaymath}
  \caption{Initial states.}
  \label{fig:init-state}
\end{figure}
Intuitively, $a_p$ is the address of the initial process, $a_r$ is
the address of the {\tt received} sequence of the initial process, and
$a_s$ is the address of the {\tt sent} sequence of the initial process.

Informally, execution of the statement initially associated with a process
may eventually (1) terminate (i.e., the statement associated with the
process becomes {\tt skip}, indicating that there is nothing left for the
process to do), (2) get stuck (i.e., the statement associated with the
process is not {\tt skip}, and the process has no enabled transitions) due
to an unsatisfied {\tt await} statement or an error (e.g., the statement
contains an expression that tries to select a component from a value that
is not a tuple, or the statement contains an expression that tries to read
the value of a non-existent field), or (3) run forever due to an infinite
loop or infinite recursion.

\standaloneOnly{}

\begin{acks}
We thank Michael Gorbovitski for supporting the use of InvTS for
automatic incrementalization of DistAlgo programs.
We are grateful to the following people for their helpful comments and
discussions: Ken Birman, Andrew Black, Jon Brandvein, Wei Chen, Ernie
Cohen, Mike Ferdman, John Field, Georges Gonthier, Leslie Lamport,
Nancy Lynch, Lambert Meertens, Stephan Merz, Don Porter, Michel
Raynal, John Reppy, Emin G\"un Sirer, Doug Smith, Gene Stark, and
Robbert van Renesse.
We thank the anonymous reviewers for their detailed and helpful
comments.
\end{acks}
\jnl{}

\def\bibdir{../../../bib}                      %
{%
\conf{
\bibliographystyle{abbrvnat}
}
\tran{}
\bibliography{\bibdir/strings,\bibdir/liu,\bibdir/IC,\bibdir/PT,\bibdir/PA,\bibdir/Lang,\bibdir/Algo,\bibdir/Perform,\bibdir/DB,\bibdir/SE,\bibdir/Sys,\bibdir/Veri,\bibdir/Sec,\bibdir/misc,\bibdir/crossref} %
}

\end{document}